\documentclass[journal]{IEEEtran}

\usepackage[english]{babel}
\usepackage[utf8]{inputenc} 
\usepackage[T1]{fontenc}    
\usepackage{hyperref}       
\usepackage{url}            
\usepackage{xcolor}
\usepackage{booktabs}       
\usepackage{amsfonts}       
\usepackage{nicefrac}       
\usepackage{microtype}      
\usepackage{multirow}
\usepackage{float}
\usepackage{array}
\usepackage{graphicx}
\usepackage[para,online,flushleft]{threeparttable}
\usepackage{cite}
\usepackage{cellspace}
\usepackage{tabstackengine}
\TABstackMath
\setstackgap{S}{0pt}

\makeatletter
\newcommand{\smallsym}[2]{#1{\mathpalette\make@small@sym{#2}}}
\newcommand{\make@small@sym}[2]{%
	\vcenter{\hbox{$\m@th\downgrade@style#1#2$}}%
}
\newcommand{\downgrade@style}[1]{%
	\ifx#1\displaystyle\scriptstyle\else
	\ifx#1\textstyle\scriptstyle\else
	\scriptscriptstyle
	\fi\fi
}

\begin{document}
	
	\title{KonVid-150k: A Dataset for No-Reference Video Quality Assessment of Videos in-the-Wild}
	
	\author{Franz~G\"otz-Hahn,
		Vlad~Hosu,~
		Hanhe~Lin,~
		and~Dietmar~Saupe~
		\thanks{F. G\"otz-Hahn, V. Hosu, H. Lin and D. Saupe are with the Department
			of Computer Science, University of Konstanz, 78464 Konstanz, Germany (e-mail: franz.hahn@uni.kn, or firstname.lastname@uni.kn).}}
	
	\markboth{Transactions on Circuits and Systems for Video Technology,~Vol.~21,~No.~2,~Feb.~2020}{}
	\maketitle
	
	\begin{abstract}
		Video quality assessment (VQA) methods focus on particular degradation types, usually artificially induced on a small set of reference videos. Hence, most traditional VQA methods under-perform in-the-wild. Deep learning approaches have had limited success due to the small size and diversity of existing VQA datasets, either artificial or authentically distorted. We introduce a new in-the-wild VQA dataset that is substantially larger and diverse: KonVid-150k. It consists of a coarsely annotated set of 153,841 videos having five quality ratings each, and 1,596 videos with a minimum of 89 ratings each. Additionally, we propose new efficient VQA approaches (MLSP-VQA) relying on multi-level spatially pooled deep-features (MLSP). They are exceptionally well suited for training at scale, compared to deep transfer learning approaches. Our best method, MLSP-VQA-FF, improves the Spearman rank-order correlation coefficient (SRCC) performance metric on the commonly used KoNViD-1k in-the-wild benchmark dataset to 0.82. It surpasses the best existing deep-learning model (0.80 SRCC) and hand-crafted feature-based method (0.78 SRCC). We further investigate how alternative approaches perform under different levels of label noise, and dataset size, showing that MLSP-VQA-FF is the overall best method for videos in-the-wild. Finally, we show that the MLSP-VQA models trained on KonVid-150k sets the new state-of-the-art for cross-test performance on KoNViD-1k, LIVE-VQC, and LIVE-Qualcomm with a 0.83, 0.75, and 0.64 SRCC, respectively. For both KoNViD-1k and LIVE-VQC this inter-dataset testing outperforms intra-dataset experiments, showing excellent generalization.
	\end{abstract}
	

	%
	\IEEEpeerreviewmaketitle

	%

	\section{Introduction}
	\label{sec:intro}
	
	\IEEEPARstart{V}{ideos} have become a central medium for business marketing~\cite{wyzowl}, with over 81\% of businesses using video as a marketing tool. Additionally, over 40\% of businesses have adopted live video formats such as Facebook Live for marketing and user connection purposes~\cite{buffer}. For consumers, video is the primary source of media entertainment; for example the average US consumer spends 38 hours per week watching video content~\cite{deloitte} and it is projected that online videos will make up more than 82\% of all consumer internet traffic by 2022~\cite{cisco2018cisco}. Streaming platforms such as YouTube report that more than a billion hours of video are watched every day~\cite{youtube}. The success of online videos is due in part to the consumer belief that traditional TV offers an inferior quality~\cite{deloitte}. Additionally, increased accessibility to video content acquisition hardware, as well as improvements in overall image quality, are a central aspect in smartphone technology advancement. Similarly, user-generated content is produced at an increasing rate, but the resulting videos often suffer from quality defects.
	
	Therefore a wide range of video producers and consumers should be able to get automated feedback on video quality. For example, user-generated video distribution platforms like YouTube or Vimeo may want to analyze new videos according to quality to separate professional from the amateur video content, instead of only indexing by video playback resolution. Additionally, with an automated video quality assessment (VQA) system, video streaming services can adjust video encoding parameters to minimize bandwidth requirements while ensuring the delivery of satisfactory video quality.
	
	A critical emerging challenge for VQA is to handle ecologically valid in-the-wild videos. In environmental psychology, ecological validity is defined as ``the applicability of the results of laboratory analogues to non-laboratory, real life settings''\cite{mckechnie1977simulation}. In our case the term can be understood as a measure for the extent to which the data represented in a dataset can be generalized to data that would be naturally encountered in the use of a technology. Concretely, this would refer to the types and degree of distortions in visual media contents of internet videos, such as those consumed on YouTube, Flickr, or Vimeo. The term in-the-wild refers to datasets that are ``not constructed and designed with research questions in mind''\cite{ang2013data}. In the case of VQA this would mean datasets that are not recorded or altered with a specific research purpose in mind, such as artificially distorting videos at variable degrees.
	
	It comes as no surprise that no-reference VQA (NR-VQA), in particular, has been a field of intensive research in the past few years achieving significant performance gains~\cite{argyropoulos2011no,chen2011prediction,valenzise2011no,saad2014blind,pandremmenou2015no,keimel2011design,zhu2014no,sogaard2015no,mittal2015completely,vega2017predictive,korhonen2018learning,korhonen2019hierarchical}. However, state-of-the-art NR-VQA algorithms perform worse on in-the-wild videos than on synthetically distorted ones. These methods aggregate individual video frame quality characteristics that are engineered for specific purposes, such as detecting particular compression artifacts. Often, these features are a balance between precision and computational efficiency. Furthermore, since there is a lack of large-scale in-the-wild video quality datasets with authentic distortions, a thorough evaluation of NR-VQA methods is difficult. Most existing databases are intended as benchmarks for the detection of those specific artificial distortions that NR-VQA algorithms have classically been designed to detect.
	
	Given the previous challenges, our first contribution is the creation of a large ecologically valid dataset, KonVid-150k. Similar to the dataset KoNViD-1k \cite{konvid}, the ecological validity of KonVid-150k stems from its size, content diversity, as well as naturally occurring, and thus representative degradations. However, being two orders of magnitude larger than existing datasets, it poses new challenges to VQA methods, requiring to train across a vast amount of content and a wide span of authentic distortions. Moreover, since a fixed budget usually constrains the development of a dataset, we needed to ensure a minimum level of annotation quality. Therefore, a part of KonVid-150k consists of 153,841 five seconds long videos that are annotated by five subjective opinions each. This set, from here on called KonVid-150k-A, is over 125 times larger than existing VQA datasets in terms of number of videos and with close to one million 
	subjective ratings over eight times larger in number of annotations~\cite{konvid,CVD2014,ghadiyaram2017capture,sinno2019large}. The dataset is accompanied by a benchmark set of nearly 1,600 videos (KonVid-150k-B) from the same source with a minimum of 89 opinion scores each. This presents a unique opportunity to analyze the trade-off between the number of training videos and the annotation noise/precision, in terms of the performance on the KonVid-150k-B benchmark dataset.
	
	This new dataset exacerbates two problems of classical NR-VQA methods. First, the computational costs of hand-crafted feature-based approaches are increased through the sheer number of videos. Second, since hand-crafted features handle in-the-wild videos worse than conventional databases, this dataset is very challenging for classical NR-VQA methods. An alternative to hand-crafted features comes with the rise of deep convolutional neural networks (DCNNs), where stacked layers of increasingly complex feature detectors are learned directly from observations of input images. These features are often relatively generic and have been proven to transfer well to similar tasks that are not too different from the source domain~\cite{gao2018blind,zhang2018unreasonable}. This suggests considering a DCNN as a feature extractor with a benefit over hand-crafted features in that the features are entirely learned from data.
	
	As a second contribution, we propose to use a new way of extracting video features by aggregating activations of all layers of DCNNs, pre-trained for classification, for a selection of frames. We adopt a strategy similar to Hosu et al.~\cite{hosu2019effective} and extract narrow multi-level spatially pooled (MLSP) features of video frames from an InceptionResNet-v2~\cite{szegedy2017inception} architecture to learn VQA. By global average pooling the outputs of inception module activation blocks, we obtain fixed sized feature representations of the frames.
	
	The third contribution of this paper consists of two network variants trained on the frame feature vectors that surpass state-of-the-art NR-VQA methods on in-the-wild datasets and train much faster than the baseline transfer learning approach of fine-tuning the entire source network. In a short ablation study we investigate the impact of architectural and hyperparameter choices of both models. Both approaches are then evaluated on existing VQA datasets consisting of authentic videos as well as those containing artificially degraded videos and show that on in-the-wild videos the proposed method outperforms classical methods based on hand-crafted features. In particular, training and testing on KoNViD-1k improves the state-of-the-art 0.80 to 0.82 SRCC. Finally, we show that training our proposed model on the new dataset of 153,841 videos with five subjective opinions each achieves a 0.83 SRCC in a cross-database test on KoNViD-1k, which outperforms state-of-the-art when training and testing on KoNViD-1k itself, which have the benefit of not being affected by any domain shift \cite{ben2007analysis}.
	
	In summary, our main contributions are:
	\begin{itemize}
		\item KonVid-150k, an ecologically valid in-the-wild video quality assessment database, two orders of magnitude larger than existing ones.
		\item The successful application of deep multi-layer spatially pooled features for video quality assessment.
		\item Three deep neural network models (MLSP-VQA-FF, -RN, and -HYB). They surpass the state-of-the-art with 0.82 SRCC versus the best existing 0.80 SRCC in an intra-dataset scenario on KoNViD-1k, and show excellent generalization in inter-dataset tests when trained on KonVid-150k, surpassing the best existing feature-based models.
	\end{itemize}
	
	
	\section{Related Work}
	
	This paper contributes to datasets and methods for video quality assessment. In this section we summarize related work in both fields as well as research in feature extraction that was influential for our work.
	
	\subsection{VQA Datasets}
	\label{sec:rw:vqa-datasets}
	
	There are a few distinguishing characteristics that divide the field of VQA datasets which are usually governed by decisions made by their creators. We will cover the characteristics differentiating the wide variety of relevant related works separately.
	
	\subsubsection{Video sources}
	The first distinguishing factor that heavily influences the use of a dataset is the source of stimuli. 
	
	The early works in the field of VQA datasets stem from 2009 to 2011. EPFL-PoliMI~\cite{de2009subjective, de2010h}, LIVE-VQA~\cite{seshadrinathan2010study, seshadrinathan2010subjective}, CSIQ~\cite{larson2010most}, VQEG-HD \cite{video2010report}, and IVP \cite{zhang2011ivp} were mostly concerned with particular compression or transmission distortions. Consequently, these early datasets contain few source videos that were degraded artifically to cover the different distortion domains. From today's standpoint the induced degradations lack ecological validity when compared to degradations observed in new videos in-the-wild. With transmission being largely an  extraneous factor, due to high-quality transmission networks, the focus of VQA datasets has been shifting towards covering a broad diversity of contents and in-the-wild distortions. 
	
	Recently designed VQA databases from 2014 to 2019 (CVD2014 \cite{CVD2014}, LIVE-Qualcomm~\cite{ghadiyaram2017capture}, KoNViD-1k~\cite{konvid}, and LIVE-VQC~\cite{sinno2019large}) have taken the first steps towards improving ecological validity. CVD2014 contains videos which were degraded with realistic video capture related artifacts. Videos in LIVE-Qualcomm, LIVE-VQC, and KoNViD-1k were either self-recorded or crawled from public domain video sharing platforms without any directed alteration of the content. 
	
	An additional side-effect of this change in dataset paradigms are differences in numbers of devices and formats represented in modern datasets. 
	\begin{itemize}
		\item CVD2014 considers videos taken by 78 different cameras with different levels of quality from low-quality camera phones to high-quality digital single-lens reflex cameras. The video sequences were captured one at a time from different scenes using different devices. They captured a total of 234 videos, three from each camera, with a mixture of in-capture distortions. While each stimulus in CVD2014 is a unique video rather than an alteration of a source video, the dataset only covers five unique scenes, which is the smallest number of unique scenes among all VQA datasets.
		\item LIVE-Qualcomm contains videos recorded using eight different mobile cameras at 54 scenes. Dominant frequently occurring distortion types such as insufficient color representation, over/under-exposure, auto-focus related distortions, blurriness, and stabilization related distortions were introduced during video capturing. In total, the 208 videos cover six types of authentic distortions, but there is no quantification as to how common these distortions are for videos in-the-wild. 
		\item LIVE-VQC contains videos captured by 80 na\"{i}ve mobile camera users, totaling 585 unique video scenes at various resolutions and orientations. 
		\item KoNViD-1k contains 1,200 unique videos sampled from YFCC100m. It is hard to quantify the number of devices covered, but in terms of content and distortion variety, it is the largest existing collection of videos. The videos in KoNViD-1k have been reproduced from Flickr, based on the highest quality download option; however, they are not the raw versions originally uploaded by users. The videos show compression artifacts, having been re-encoded to reduce bandwidth requirements. 
	\end{itemize}
	We are employing a strategy similar to KoNViD-1k, however we obtained the originally uploaded versions of the videos to re-encode them at a higher quality. We aim to reduce the number of encoding artifacts while keeping the file size manageable for distribution in a crowdsourcing study with an average of 1.23 megabytes per video.
	
	\subsubsection{Subjective assessment}
	The second distinguishing factor is the choice of subjective assessment environment. VQA has been a field of research since before the time when video could easily and reliably be transmitted over the Internet. Consequently, early datasets have all been annotated by participants in a lab environment. This allows for assessment of quality under strictly-controlled conditions with reliable raters, giving an upper bound to discriminability. With dataset sizes increasing, due to a push for more content diversity and transmission rates improving, crowdsourcing has become an affordable and fast way of annotating multimedia datasets with subjective opinions. In a lab setup it is practically infeasible to handle annotation of tens of thousands of items. The downside of crowdsourcing is a reduced level of control over the environment, resulting in potentially lower quality of annotation. However, with careful quality control considerations a crowdsourcing setup can achieve an annotation quality comparable to lab setups\cite{saupe2016crowd}. Concretely, CVD2014 and LIVE-Qualcomm are annotated in a lab environment, while KoNViD-1k and LIVE-VQC are both annotated using crowdsourcing. Considering the sheer size of our dataset, we also employed a crowdsourcing campaign with rigorous quality control in the form of an initial quiz and interspersed test questions to ensure a good annotation quality.
	
	\subsubsection{Number of observers}
	A third factor that has been studied only very little thus far is the choice of numbers of ratings per video. With a few exceptions, early works in lab environments ensured at least 25 raters per stimulus. Additionally, it has been a common approach that all participants rated all stimuli. 
	
	Recent works\cite{sinno2019large} have increased the number of ratings per stimulus to above 200 to ensure very high quality annotation. However, given a fixed, affordable budget of annotations, one must consider the trade-off between the benefit of slightly more accurate quality scores for a small number of stimuli and the potential increase in generalizability when annotating more stimuli with fewer votes. The 8-fold increase in numbers of ratings per stimulus when going from the generally accepted 25 to 200 ratings could just as well be invested in an 8-fold increase of numbers of stimuli, each rated 25 times. The increase of the precision of the experimental MOS suffers from diminishing returns as the number of raters increases. Since the precision gain per vote is highest at none or few ratings, careful considerations have to be made with respect to the distribution of annotation budgets across an unlabeled dataset. This is especially true in the wake of deep learning approaches outperforming classical methods in many computer vision tasks, as deep learning models are known to be robust to noisy labels~\cite{rolnick2017deep} but also hungry for input data.
	
\begin{figure}[t!]
		\centering
		\includegraphics[width=\columnwidth]{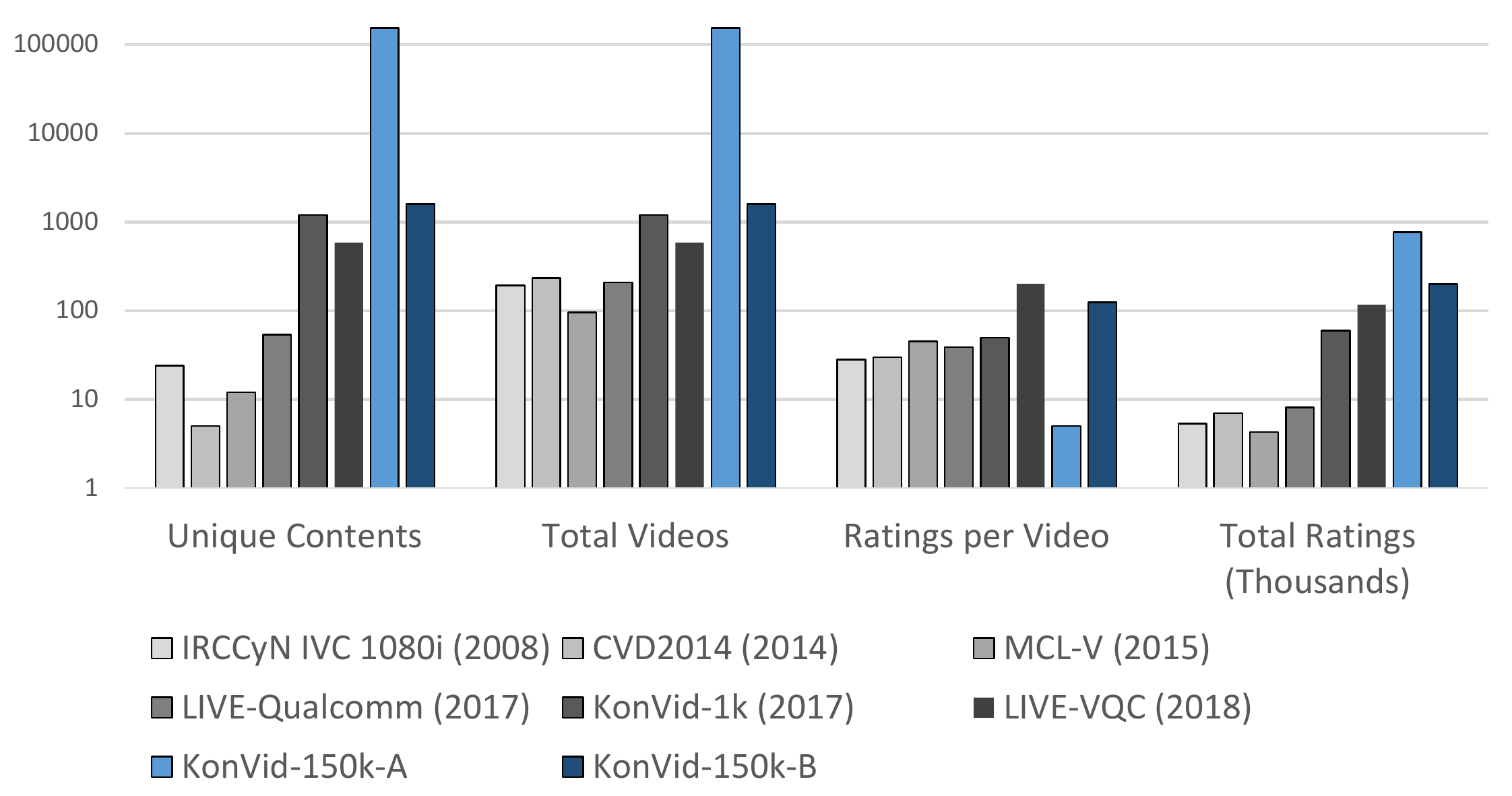} \\
		\caption{Comparison of size characteristics of current VQA  datasets. Our proposed datasets, KonVid-150k-A and KonVid-150k-B are represented by the two right most bars of the histograms. Note the logarithmic scale.}
		\label{fig:rw:datasets}
	\end{figure}

	Figure~\ref{fig:rw:datasets} shows a comparison of relevant VQA datasets on some of these characteristics. 
	There is an evident progression to a wider variety of contents in the last few years. We are attempting to push this boundary much further by exploring the trade-off between the number of ratings per video and the total annotated stimuli.
	
	\subsection{Feature Extraction} 
	
	There have been several recent works that inspired our approach for feature extraction. The BLINDER framework \cite{gao2018blind} was an initial work that utilized multi-level deep-features to predict image quality. They resized images to $224 \times 224$ and extracted a feature vector from each layer of a pre-trained VGG-net. Each of these features vectors was then fed into separate SVR heads and trained, such that the average layer-wise scores predict the quality of an image. BLINDER was evaluated on a variety of IQA datasets, vastly improving the state-of-the-art. \cite{hosu2019effective} went a step further by utilizing deeper architectures to extract features, such as Inception-v3 and InceptionResNet-v2. Furthermore, features were aggregated from multiple levels and extracted from images at their original size. This retained detailed information that would have been lost by down-sizing the inputs. Moreover, it allowed linking information coming from early levels (image dependent) and general category-related information from the latter levels in the network.
	
	We use the same approach as presented in \cite{hosu2019effective} to extract sets of features of video frames. The layers of the DNNs are a basic measure for the level of complexity that the feature can represent. For example, first layer features resemble Gabor filters or color blobs, while features in higher levels correspond to semantic entities such as circular objects with a particular texture or even faces. Changes in the response of different features can, therefore, encode temporal information. For example, it is reasonable to assume that a change in the overall response of low-level Gabor-like features can indicate the rapid movement of an object. Consequently, learning from frame-level features allows to learn the effect of temporal degradations on video quality indirectly.
	
	In~\cite{varga2020multi} a similar approach was used for the purpose of NR-VQA. The method extracted features for intra-frames, averaging them along the temporal domain to obtain a video-level feature vector. The final video quality prediction is done by an SVR. In our approach we go beyond this by considering both an average feature vector with our MLSP-VQA-FF architecture, as well as an LSTM model that takes a set of consecutive features of frames as input, leveraging temporal information of feature activations.
	
	\subsection{NR-VQA} 
	
	Existing NR-VQA methods can be differentiated based on whether they are based solely on spatial image-level features or also explicitly account for temporal information. In general, however, all recently developed models are learning-based.
	
	Image-based NR-VQA methods are mostly based on theories of human perception, with natural scene statistics (NSS)~\cite{srivastava2003advances} being the predominant hypothesis used in several works, such as the naturalness image quality evaluator (NIQE)~\cite{mittal2012making}, blind/referenceless image spatial quality evaluator (BRISQUE)~\cite{mittal2012no}, feature-map-based referenceless image quality evaluation engine (FRIQUEE)~\cite{xu2014no} and high dynamic-range image gradient-based evaluator (HIGRADE)~\cite{kundu2017no}.  NSS hypothesizes that certain statistical distributions govern how the human visual system processes particular characteristics of natural images. Image quality can be derived by measuring the perturbations of these statistics. The approaches above have been extended to videos by evaluating them on a representative sample of frames and aggregating the features by averaging.
	
	Approaches that consider temporal features, so-called general-purpose VQA methods, are less numerous and more particular in their approach. In~\cite{saad2014blind}, the authors extended an image-based metric by incorporating time-frequency characteristics and temporal motion information of a given video using a motion coherence tensor that summarizes the predominant motion directions over local neighborhoods. The resulting approach, coined V-BLIINDS, has been the de facto standard that new NR-VQA methods are compared with.
	
	Apart from V-BLIINDS, several other machine-learning-based models for NR-VQA have been proposed. Regrettably, most have only been evaluated on older datasets such as LIVE-VQA, making comparisons across multiple datasets difficult. Moreover, their codes are not publicly available, further exacerbating this issue. The three most notable examples are the following. V-CORNIA~\cite{xu2014no} is an unsupervised frame-base feature-learning approach that uses Support Vector Regression (SVR) to predict frame-level quality. Temporal pooling is then applied to obtain the final video quality. SACONVA~\cite{li2016no} extracts feature descriptors using a 3D shearlet transform of multiple frames of a video, which are then passed to a 1D CNN to extract spatio-temporal quality features. COME~\cite{wang2018come} separated the problem of extracting spatio-temporal quality features into two parts. By fine-tuning AlexNet on the CSIQ dataset, spatial quality features are extracted for each frame by both max pooling and computing the standard deviation of activations in the last layer. Additionally, temporal quality features are extracted as standard deviations of motion vectors in the video. Then, two SVR models are used in conjunction with a Bayes classifier to predict the quality score.
	
	The state-of-the-art in blind VQA is set by two recently published approaches, namely TLVQM~\cite{korhonen2019hierarchical} and 3D-CNN + LSTM~\cite{you2019deep}. The former is a hierarchical approach for feature extraction. It computes two types of features: low complexity features characterizing temporal aspects of the video for all video frames, and high complexity features representing spatial aspects. High complexity features relating to spatial activity, exposure, or sharpness,  are extracted from a small representative subset of frames. TLVQM achieves the best performance on LIVE-Qualcomm and CVD2014. The latter is an end-to-end DNN approach, where 32 groups of 16 224$\times$224 crops of frames are extracted from the original video and individually fed into a 3D-CNN architecture that outputs a scalar frame-group quality. This is then subsequently passed to an LSTM that predicts the overall video quality. This approach sets the state-of-the-art for KoNViD-1k, besting TLVQM slightly. 
	
	There has been a body of work by another author on NR-VQA \cite{varga2019no, varga2019no2, varga2020multi}. However, there are concerns about the validity of the published performance values~\cite{gotz2020critical}. Specifically, it has been shown that the performance values reported in both \cite{varga2019no} and \cite{varga2019no2} were obtained with implementations containing some forms of data leakage. In both cases, the fine-tuning stage of the two-stage process embedded information about the test sets into the model used for feature extraction. Furthermore, in \cite{gotz2020critical} it was shown that fine-tuning prior to feature extraction had much less impact on the final performance than claimed. Since \cite{varga2020multi} is using a similar two-stage approach involving fine-tuning and feature extraction, and there is a substantial improvement in performance from the non-fine-tuned to the fine-tuned implementation, we hold some reservations as to the validity of the reported performance values.
	
	\section{Dataset Implementation Details}
	
	In this section, we introduce the video dataset in two parts. First, we discuss the design choices and gathering of the data in Section~\ref{sec:db:dataset} alongside an evaluation of the diversity captured by the dataset in relation to existing work in Section~\ref{sec:db:eval}. Then, Section~\ref{sec:db:annotation} follows up with details regarding the crowdsourcing experiment to annotate the dataset. 
	
	\subsection{Video Dataset}
	\label{sec:db:dataset}
	
	Our main objective was to create a video dataset that covers a wide variety of contents and quality-levels as commonly available on video sharing websites. For this reason, we took a similar approach to collect our data as was done for KoNViD-1k, with an additional step to improve the quality of the videos. In KoNViD-1k all collected videos had been transcoded by Flickr, to reduce their bandwidth requirements and standardizing them for playback. Consequently, noticeable degradation was introduced relative to the original uploads. Flickr allows the uploading of video files of most codec and container combinations, resolutions, and durations. However, they re-encode the uploaded videos to common resolutions such as HD, Full HD, strongly compressing them. 
	
	The Flickr API allows access to metadata that links to the original, raw uploads. As these raw uploads are often very large and come in many different formats, they cannot directly be used for crowdsourcing. Therefore, we proceeded as follows. We downloaded authentic raw videos that had an aspect ratio of 16:9 and resolution higher than 960$\times$540 pixels. Then we rescaled them to 960$\times$540, if necessary, and extracted the middle five seconds. Finally, we re-encoded them using FFmpeg at a constant rate factor of 23, which balances visual quality and file size. The resulting files have an average size of 1.23 megabytes.
	
\begin{figure}[t]
		\centering
		\includegraphics[width=0.49\textwidth]{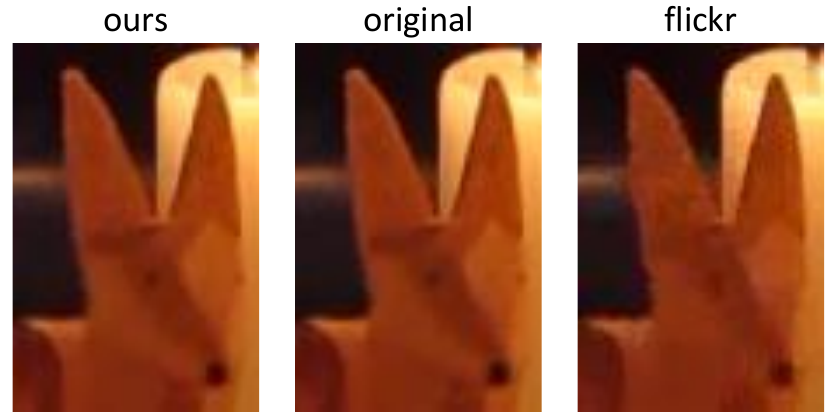} \\
		\caption{Comparison of the quality of the original (center) to the version Flickr provides (right) and our transcoded version (left).}
		\label{fig:db:flickr-ours-comparison}
	\end{figure}

	Figure~\ref{fig:db:flickr-ours-comparison} is a visual comparison of the differences, showing a small crop of a frame of the originally uploaded video together with the two re-encodings offered by Flickr and our own version. Compression artifacts are clearly visible in the Flickr re-encoded version, whereas our re-encoding is very similar to the original.
	
	For each video, we extracted meta-information that identifies the original encoding, including the codec and the bit-rate. Furthermore, we collected social-network attributes such as the number of views and likes and publication dates that indicate the popularity of videos. In total, this collection amounts to 153,841 videos. We believe that all the additional measures we have taken to refine our dataset significantly improved its ecological validity, and thus the performance of VQA methods trained on it in the future.
	
	\subsection{Dataset Evaluation}
	\label{sec:db:eval}
	
	In order to evaluate the diversity of KonVid-150k, which is our main objective with this dataset, we will now demonstrate that it is not only the largest annotated VQA dataset in terms of video items, but also the most diverse in terms of content. First, we need a measure for content diversity. For this purpose we extract the activations of the last fully-connected layer of an Inception-ResNet-v2 model pre-trained on ImageNet for each frame. To represent a given video, we average these activations over all frames to obtain a 1792-dimensional content feature. A similar approach has been used in the image quality domain before to create a subset of data that is diverse in content~\cite{hosu2020koniq}. 
	
	\begin{figure*}[t!]
		\centering
		\includegraphics[width=0.99\textwidth]{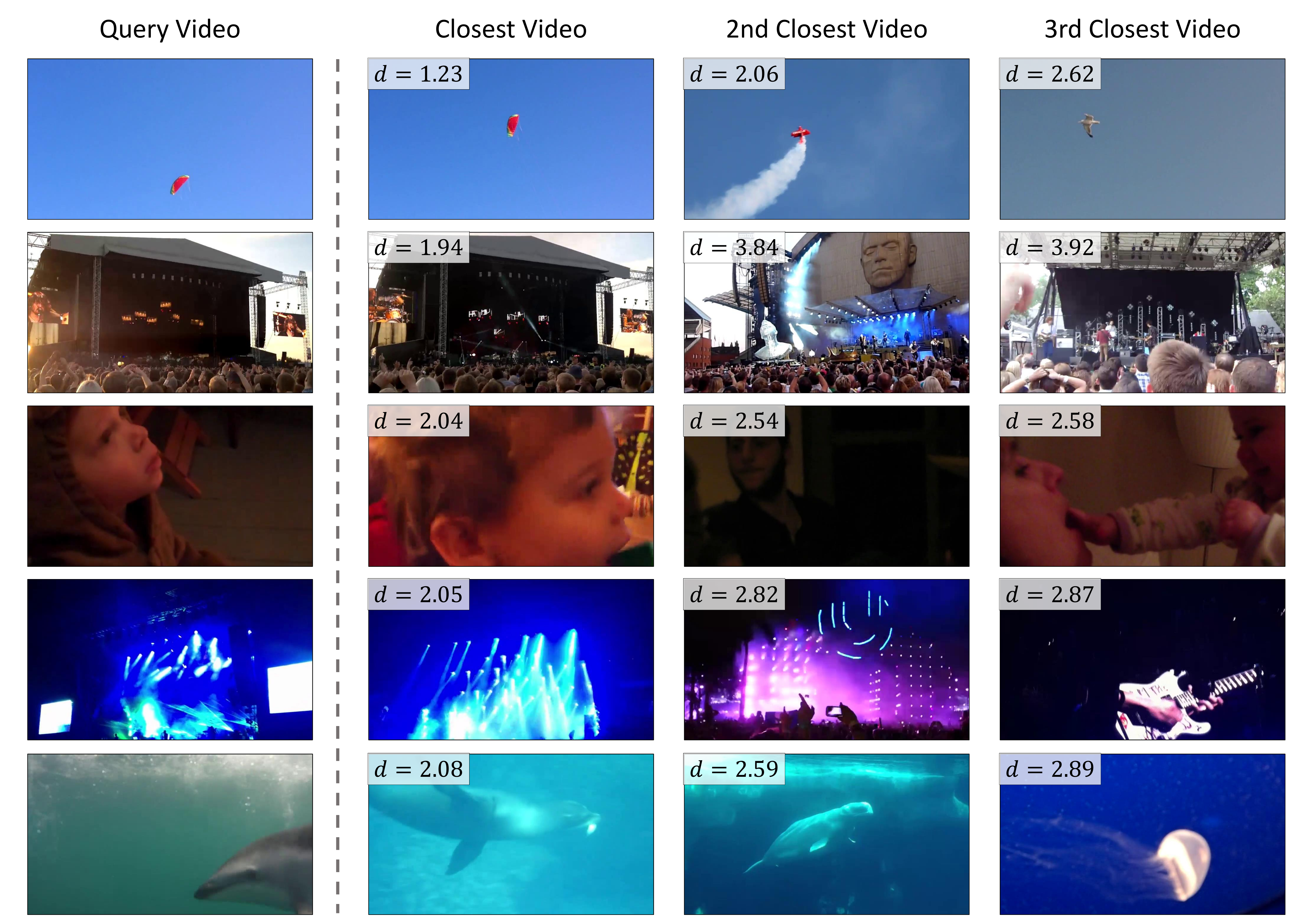} \\
		\caption{Still images from videos closest to the query video on the left as measured by the Euclidean distance $d$ in the feature space of top-layer features from Inception-ResNet-v2. This shows the utility of activations of layers from pre-trained DCNNs for usage in a content similarity measure. Even though only the 1792 activations of the last layer were used, which are commonly understood to focus on semantic entities more so than low level structures, these features encode useful information.}
		\label{fig:db:content-distance}
	\end{figure*} 
	
	Figure~\ref{fig:db:content-distance} is an illustration of the usefulness of these content features to assess content similarity. Given a query video taken from KoNViD-1k on the left we compute the Euclidean distance in content feature space to all other videos in the dataset. On the right we show still frames from the three videos with smallest distance to the query. We can see that close proximity in content feature space seems to correspond to semantically similar video content. The images in the first row show flying objects in a blue sky, where the color of the object as well as the color of the sky seem to influence the distance in content feature space. In the second row we can see that crowds in front of a stage are located in close proximity in content feature space. Images in the third row show that videos containing heads, but especially babies are encoded similarly in the 1792-d content feature vectors. Light shows and underwater videos, as seen in the fourth and fifth rows, can also be retrieved by querying nearest neighbours of an appropriate video. It is to be noted that the closest videos for rows one, two and four are near duplicates. The recordings seem to be from different periods of time of the same scene.
	
	Therefore, the extracted features are useful as an information retrieval tool, and we make use of it to quantify the degree by which a video dataset covers the content of competing datasets. For this purpose we represent a video dataset by its corresponding set of content feature vectors, $X = \{x_i~|~i=1,...,N\}$, where $N$ is the number of videos in the dataset. We consider the Euclidean distance of a point $x$ in feature space to a (finite) point set $Y$, $d(x,Y)=\min \{d(x,y)~|~y \in Y\}$. For two finite point sets $X=\{x_1,...,x_n\}$, $Y=\{y_1,...,y_m\}$ and any given distance $s \ge 0$, we define the fraction or ratio of the first dataset $X$, that is covered by the dataset $Y$ at distance $s$ as 
	$$
	C_{Y,s}(X) = \frac {|\{x \in X~|~ d(x,Y) \le s \}|}{|X|}
	$$ 
	where $|A|$ denotes the cardinality of a set $A$. For example, if $X \subseteq Y$, then $Y$ covers $X$ perfectly at distance zero, i.e., $C_{Y,0}(X) = 1$. Or, if $C_{Y,1}(X) = 0.8$, then this means that the union of all balls of radius 1 centered at the points of the set $Y$ contain 80\% of the points in $X$. The function $s \mapsto C_{Y,s}(X)$ thus comprises the cumulative histogram of the individual distances $d(x,Y)$ for all $x \in X$. 
	
	When comparing the coverage two datasets with respect to each other, we check the corresponding cumulative histograms showing the coverage of one dataset by the other. The dataset with the topmost cumulative histogram then can be considered to be the dominant one that covers the competing one.
	
\begin{figure}[t!]
		\centering
		\includegraphics[width=0.5\textwidth]{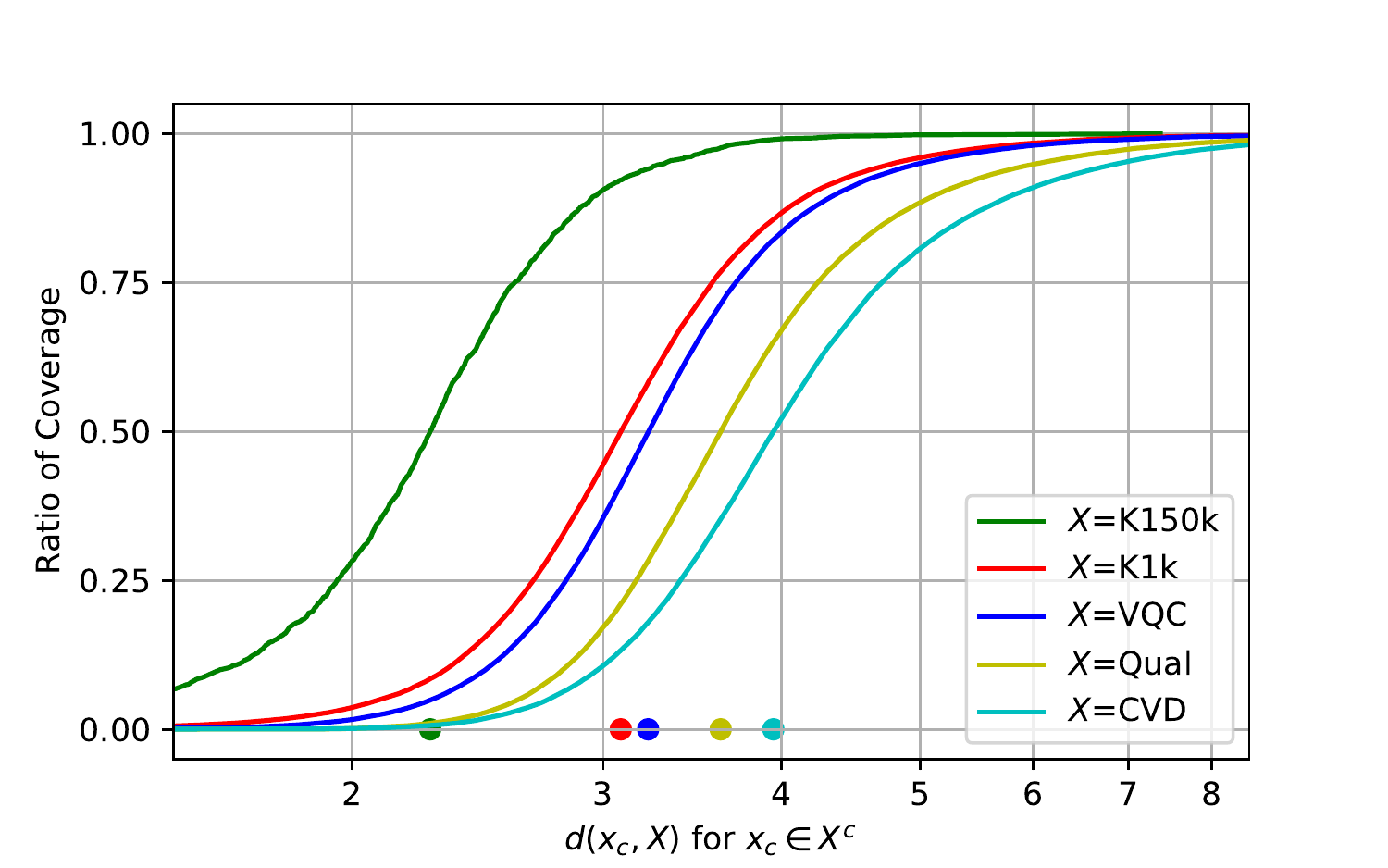} \\
		\caption{This figure shows how well a video dataset covers all others together. The curves are the empirical cumulative histograms of Euclidean distances $d(x_c,X)$ for all $x_c\in X^c$, where $X^c$ is the complement to $X$, i.e., the union of the other datasets. The green, red, blue, yellow, and cyan lines refer to $X$ being KonVid-150k, KoNViD-1k, VQC, Qualcomm, and CVD 2014, respectively. KonVid-150k covers the other datasets the best, as the green plot has the largest area under the curve and it has the smallest median distance of approximately 2.3 at coverage ratio 0.5. This means that for half of the videos in all other datasets, there is a similar video in KonVid-150k that has a distance in content feature space of at most 2.3.}
		\label{fig:db:emp_cum_density_onevall}
	\end{figure}

	To compare the diversity of content for several given datasets $X_1,\ldots,X_K$, let us form their union $Z = X_1 \cup \cdots \cup,X_k$ and consider how well each dataset $X_k$ covers all the others, i.e., the complement $X_k^c = Z \backslash X_k$. For this purpose we compute the cumulative histograms $C_{X_k,s}(X_k^c)$ for $k = 1,\ldots,K$. Figure~\ref{fig:db:emp_cum_density_onevall} shows the result for the five datasets KonVid-150k, KoNViD-1k, VQC, Qualcomm, and CVD 2014. Here, KonVid-150k clearly has the best coverage of contents present in the other datasets, as it has the largest area under the curve.
	
	To summarize the coverage of one dataset $X$ by another, $Y$, by a single number rather than the curves of the cumulative histogram of distances, we define the one-sided distance of $X$ from $Y$ as 
	$$
	d(X,Y) = f(d(x_1,Y), d(x_2,Y), ..., d(x_n,Y))
	$$ 
	where $f$ is a scalar, non-negative function. For example, if $f$ is the maximum function, then $d(X,Y)$ is known as the one-sided Hausdorff distance. For our purpose, the median is better suited as it is less sensitive to outliers. The distance $d(X,Y)$ can be understood as a simplified indicator for the coverage of $X$ by $Y$. These medians are shown in Figure~\ref{fig:db:emp_cum_density_onevall} by the bullet dots at the coverage ratio of 0.5.
	
	\begin{figure}[t!]
		\centering
		\includegraphics[width=0.5\textwidth]{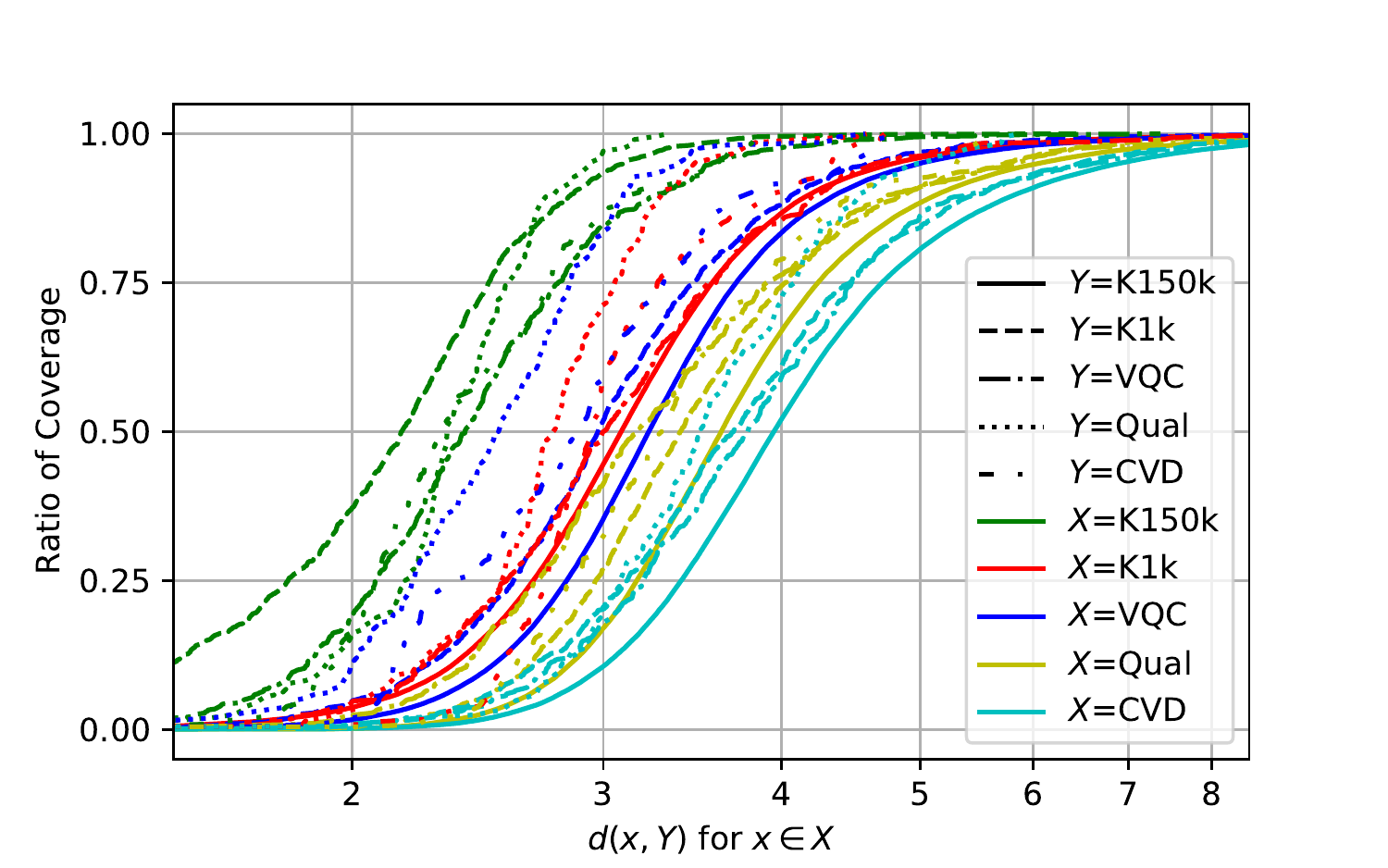} \\
		\caption{Pairwise comparison of content coverage. Empirical cumulative histograms of $d(x,Y)$ for all $x \in X$. The green, red, blue, yellow, and cyan line colors refer to the covering set $Y$ and the different line styles refer to $X$ being KonVid-150k, KoNViD-1k, CVD 2014, Qualcomm, and VQC, respectively. As expected from the previous figure, KonVid-150k covers the other datasets the best, indicated by the four green plots consistently falling to the left of their counterparts. The summarizing statistics, $d(X,Y)$ can be taken from the intersections of the graphs with  the }
		\label{fig:db:emp_cum_density}
	\end{figure} 
	
	Figure~\ref{fig:db:emp_cum_density} then shows $d(X,Y)$ for the competing dataset pairs individually. It can be seen that KonVid-150k covers the contents of competing datasets the best, as the green curves are strictly above the cumulative histograms for the other datasets. Moreover, the other datasets cover the content space of KonVid-150k the worst, as the solid lines depicting the coverage of KoNViD-1k, CVD 2014, Qualcomm, and VQC of KonVid-150k are generally to the right of the other three for the respective dataset.
	
	These findings are an indication that our proposed dataset KonVid-150k is comprised of a large variety of contents with good coverage of the contents contained in existing works.
	
	\subsection{Video Annotation}
	\label{sec:db:annotation}
	\begin{figure*}[t!]
		\centering
		\includegraphics[width=0.85\textwidth]{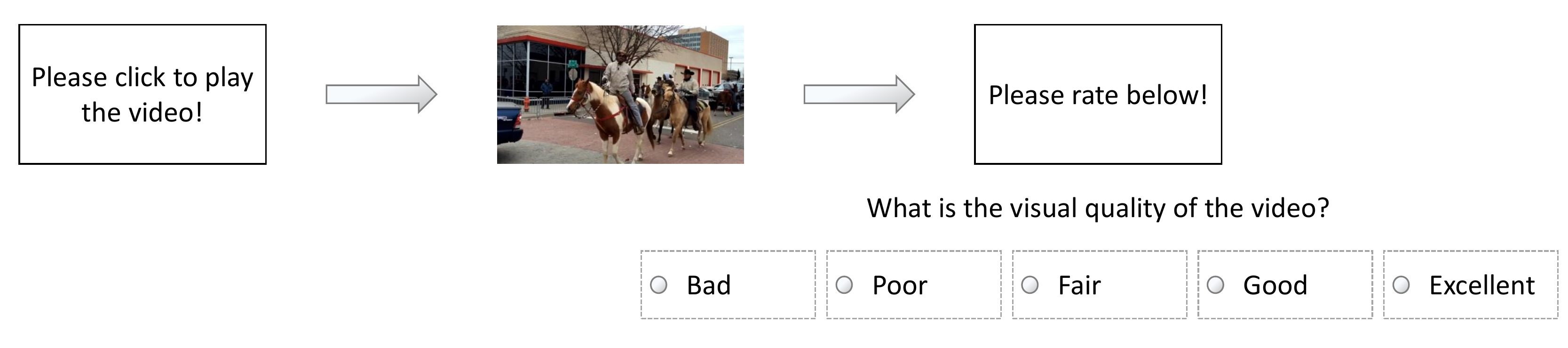} ~ 
		\caption{Illustration of the crowdsourcing video playback workflow. A worker is first presented with a white box of 960x540 pixels. Upon clicking the box, the video plays in its place. Playback controls are disabled and hidden. Upon finishing, the video is hidden and replaced with a white box that informs the participant to rate the quality on the Absolute Category Rating (ACR) scale shown below. The rating scale is only shown upon completion of video playback.} 
		\label{fig:db:crowd-setup}
	\end{figure*}
	
	We annotated all 153,841 videos for quality in a crowdsourced setting on Figure Eight\footnote{http://www.figure-eight.com/ (now https://appen.com/)}. First, each participant was presented with instructions according to VQEG recommendations~\cite{tutorial2004objective}, which were modified to our requirements. Here, participants were introduced to the task and provided with information about types of degradation, e.g., poor levels of detail, inconsistencies in color and brightness, or imperfections in motion. Next, we provided examples of videos of a variety of quality levels with a brief description of identifiable flaws and instructed the reader on the workflow of rating videos, which is illustrated in Figure~\ref{fig:db:crowd-setup}. Finally, we informed participants about ongoing hidden test questions that were presented throughout the experiment, as well as the minimum resolution requirement that enabled them to continue participating in the experiment. This was checked before the playback of any video.
	
	During the actual annotation procedure, for each stimulus, workers were first presented with a white-box of the size of the video that also functioned as a play button. Then, the video was shown in its place with the playback controls hidden and deactivated. After playback finished, it was hidden, and the rating scale was revealed below it. This setup ensured that neither the first nor the last still frame of the video were influencing the worker's rating, and no preemptive rating could be performed before the entirety of the video had been seen. An option to replay the video was not provided so as to improve attentiveness and ensure that the obtained score is the intuitive response from the worker. Additionally, playback of any other video on the page was disabled until the currently playing video was finished, in order to better control viewing behavior and discourage unreliable or random answers.
	
	According to Figure Eight's design concept, crowd workers submit batches of multiple ratings in so-called pages. Each page has a fixed batch size of rows, where each row conventionally represents a single item. Due to constraints on the number of rows allowed per study, we grouped 15 stimuli by random selection into each row, with a page size of ten rows per page, totaling to 150 videos per batch, respectively page. 
	
	Moreover, the design concept intends a two-stage testing process, where workers are first presented with a quiz of test questions followed by subsequent pages where test questions are randomly inserted into the data acquisition process. Test questions are not distinguishable from conventional annotation items.
	
\begin{figure}[t!]
		\centering
		\includegraphics[width=0.35\textwidth]{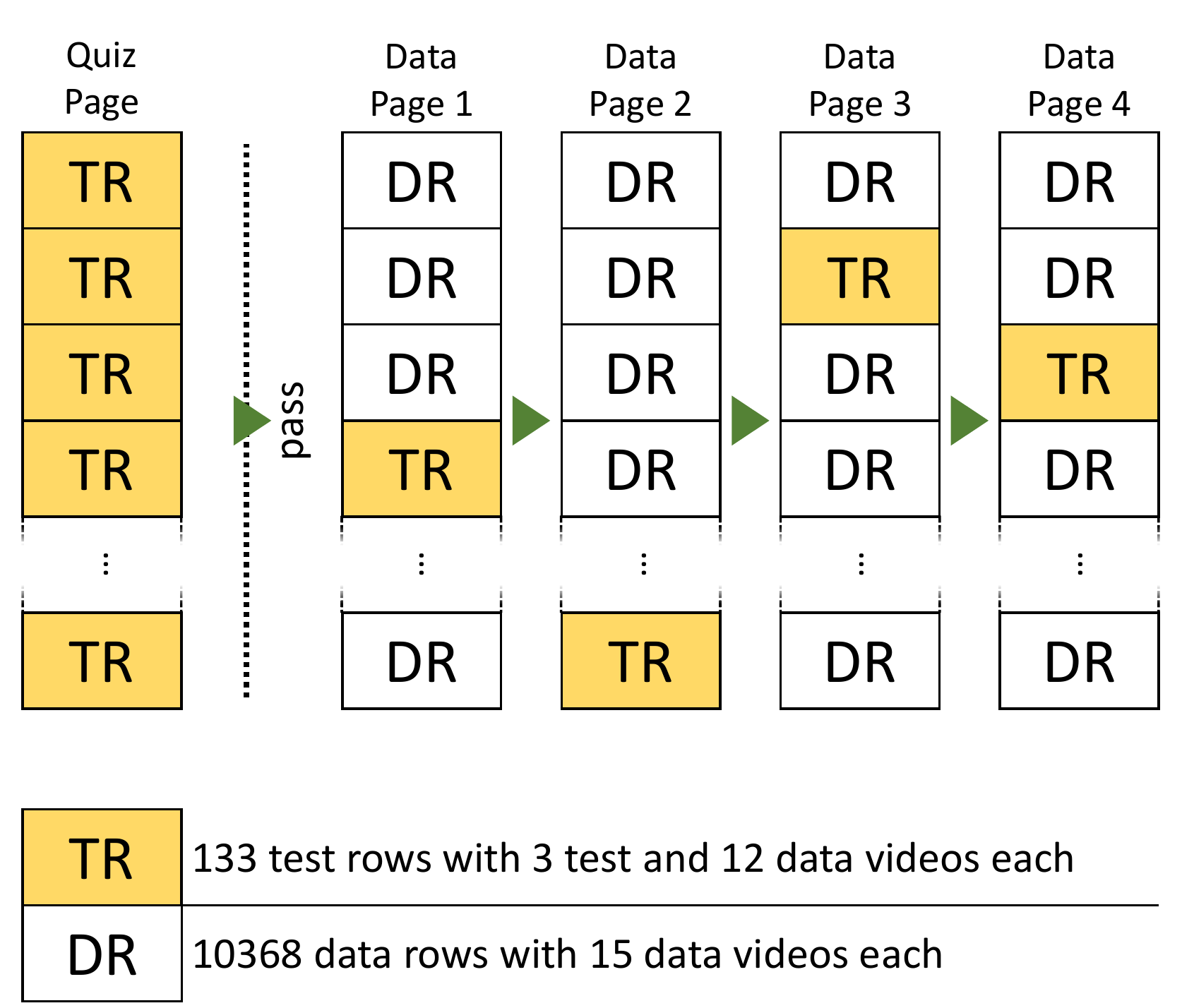} ~ 
		\caption{Simplified work flow diagram of the experiment. A worker is first presented with a quiz page of test rows (TR, in yellow) with three test videos and twelve data videos each. Upon passing the quiz with $\geq70\%$ accuracy they proceed to answer data pages with one test row per page. Data rows (DR, in white) contain 15 data videos. Data rows are annotated by five unique participants. Test rows can be answered once by each worker.}\label{fig:db:figure-eight-flow}
	\end{figure}

	In our implementation, illustrated in Figure~\ref{fig:db:figure-eight-flow}, we interspersed three test videos with twelve videos randomly sampled from the dataset in each row with test questions. The test videos were sampled from hand-picked set of videos, which in one part was made up of very high-quality videos obtained from Pixabay\footnote{http://pixabay.com} and in another of heavily degraded versions of them. Therefore, we defined the ground truth quality of each test video as either excellent or bad, respectively. We performed a confirmation study to ensure that the perceived quality of these videos was rated at the very top or bottom ends of the 5-point ACR scale. 
	
	In the second stage, after the quiz, consisting of only test rows, workers annotated 150 videos in 10 rows per page. On each page, we included one further test row at a random position.
	
	Participants had to retain at least 70\% accuracy on test questions throughout the experiment. Data entered from workers that dropped below this threshold were removed from our study, and the corresponding videos were scheduled for re-annotation. 
	
	
	When running a study on Figure Eight, the experimenter decides the number of ratings per data row, as well as the pay per page. The latter was set such that with eight seconds per video, including five seconds for viewing and three seconds for making the decision, a worker would be paid USD 3 per hour. We had compiled 10,368 data rows of 15 data videos each. These data rows were presented to five workers each, yielding 155,520 annotated video clips. From these, 152,265 were valid\footnote{\label{note2}In some rare ($\leq1\%$) cases users bypassed our restrictions by disabling javascript and were able to proceed without actually rating the videos. In that case the required 5 votes were not met, and we had to discard this video. Additionally, not all videos were readable by the Python libraries we used as feature extractors. Those videos were also removed.} and were retained, forming our larger dataset, called KonVid-150k-A.
	
	Each of the 10,368 data rows was presented to five workers. There were altogether 133 test rows for presentation to all crowd workers. However, each crowd worker could annotate any given test row at most once. Since 12 of the 15 videos in a test row were sampled from the set of data videos, we thus obtained far more than five ratings for each of these individual videos. In total, 1,596 data videos were used in the 133 test rows and were rated between 89 and 175 times, due to randomness in test question distribution. We separated 1,575 valid$^{\ref{note2}}$ videos of this very extensively annotated set in a new dataset and call it KonVid-150k-B. As a random subset of the entirety of our videos selected from Flickr, it is ecologically valid and from the same domain as the other data videos. This dataset will be used as a test set for the evaluation of our models trained on KonVid-150k-A.
	
	The choice for five individual ratings per data row was based on a small scale pilot study with a subset of 600 randomly sampled videos. For this subset we obtained two sets of 50 opinion scores for each video with a similar experimental setup as described above. We then evaluated the SRCC between a MOS comprised of a random sample of $n$ votes from one set to the MOS of the other set. At 5 votes this SRCC reached 0.8, which we considered to be a good threshold. For reference, the SRCC between the two independent samplings of 50 votes settled at 0.9. Further investigation of the feasibility of our choice of 5 ratings is contained in more detail in Section \ref{sec:modeleval:trainingschemes}.
	
\begin{figure}[t!]
		\centering
		\includegraphics[width=0.45\textwidth]{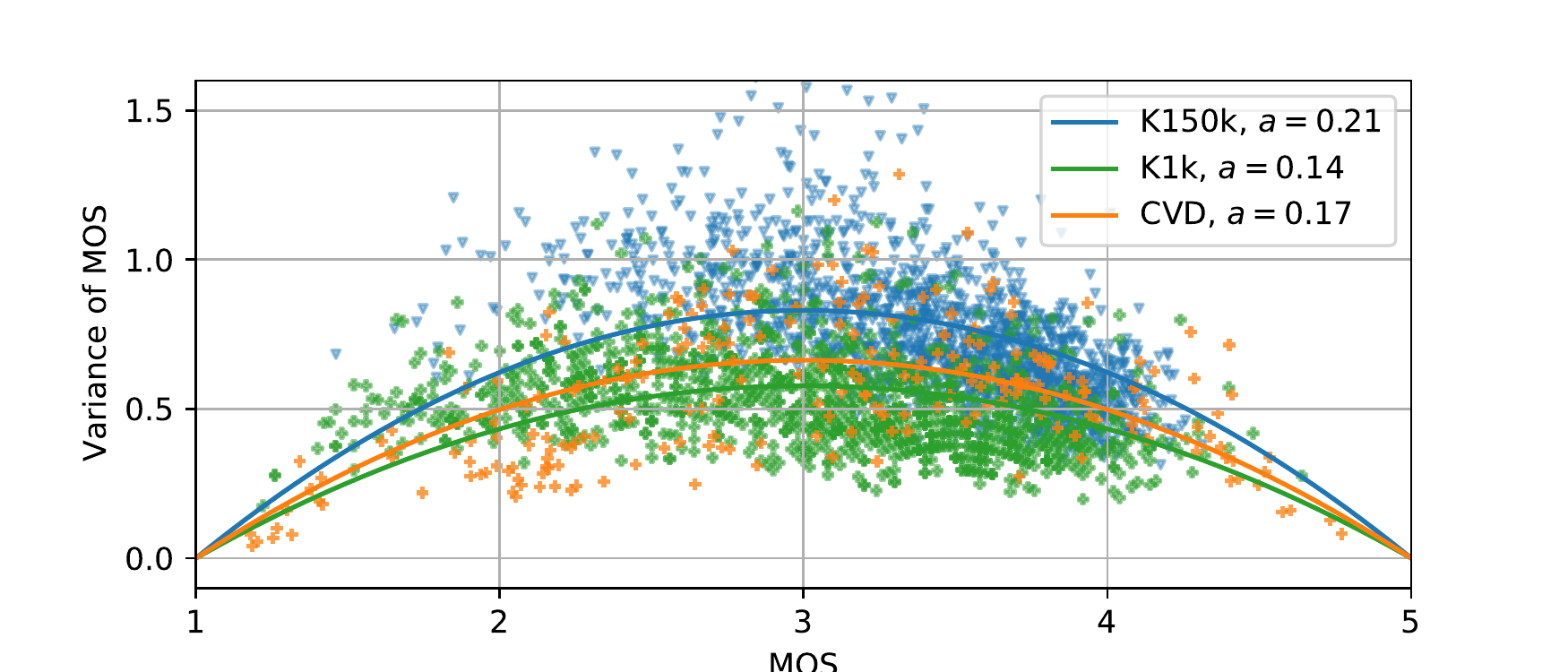} ~ 
		\caption{Comparison of the SOS hypothesis~\cite{hossfeld2011sos} of KoNViD-1k, CVD2014, and KonVid-150k-B. The SOS parameter for the three datasets are $a=0.14$, $a=0.17$, and $a=0.21$, respectively. For VQA the recommended range is $a \in [0.11,0.21]$, which shows that KonVid-150k is of sufficient annotation quality.}\label{fig:db:sosmos}
	\end{figure}

	Another common characteristic to compare the annotation quality of different studies is by evaluating the standard deviation of opinion scores (SOS) as a function of MOS. It follows the basic idea that in experimental studies that are conducted in a quality controlled manner subjective opinions will vary only to a certain extent, as the experimental setup ensures similar test conditions. In the case of the 5-point scale we used in our experimental setup, the maximum SOS is found at a MOS of 3, while the minimum will always be at the extremes of the rating scale. However, computing the average SOS over all videos is not an unbiased indicator, as datasets commonly do not contain a uniform distribution of videos in relation to the MOS. Instead, the variance $\sigma^2$ is modelled as a quadratic function of the MOS~\cite{hossfeld2011sos}, which in the case of a 5-point scale is described as:
	
	\begin{equation}
		\textrm{SOS}(\textrm{MOS})^2 = a(-x^2+6x-5),
	\end{equation}
	
	where the SOS parameter $a$ better indicates the variance of subjective opinions for any particular experimental study. Moreover, it has been shown to correlate with task difficulty~\cite{janowski2015accuracy} and can be used to characterise application categories. A reasonable range for the SOS parameter in the domain of VQA has been reported to be $a \in [0.11, 0.21]$, with $a_{\textrm{KoNViD-1k}}=0.14$ and $a_{\textrm{CVD2014}}=0.17$. In the case of LIVE-Qualcomm and LIVE-VQC, no SOS parameter has been reported and the publicly available annotation data does not allow for such an analysis, as only the MOS values for videos in these specific datasets are available. We have evaluated the SOS hypothesis for KonVid-150k as well, however we have limited it to the KonVid-150k-B set, as the discretized MOS values for the larger KonVid-150k-A set render it incompatible to the other datasets. Nonetheless, KonVid-150k-B is a good estimation of what can be expected in terms of annotation quality of KonVid-150k as a whole. Figure~\ref{fig:db:sosmos} shows the comparison between KoNViD-1k, CVD2014, and KonVid-150k-B, where the latter has an SOS parameter of $a_{\textrm{KonVid-150k-B}}=0.21$, which lies within the recommended range for VQA experiments.
	
	\section{Video Quality Prediction}
	
	The na\"ive way to perform transfer learning for tasks related to visual features with small sets of data is removing the head of a pre-trained base-model and replacing it with a small fully connected head. By freezing the layers in the base-model it's predictive power can be used to perform well on the new task. After training this new header, it is not uncommon to unfreeze all layers and fine-tuning the entire trained network with a low learning rate to improve predictive power even more. However, this approach has three important downsides.
	
	\begin{enumerate}
		\item First, the new task is trained based on the highest level features in the base-model. These features are particularly tuned to detecting high-level semantic features that are useful in the detection of objects present in the image. However, for tasks such as quality, low-level features with a small receptive field are arguably more important.
		\item Secondly, for each forward and backward pass the entire base-model has to be present in memory, which contain many more weights than the header network that is being trained. Consequently, training is slowed down a lot.
		\item Finally, the last fine-tuning step is prone to overfitting, as the high capacity of the base-model alone allows the network to memorize training data rather than extracting useful general features. Careful hyperparameter tuning is therefore required, to ensure this step is successful in improving performance.
	\end{enumerate}
	
\begin{figure*}[t]
		\centering
		\includegraphics[width=0.95\textwidth]{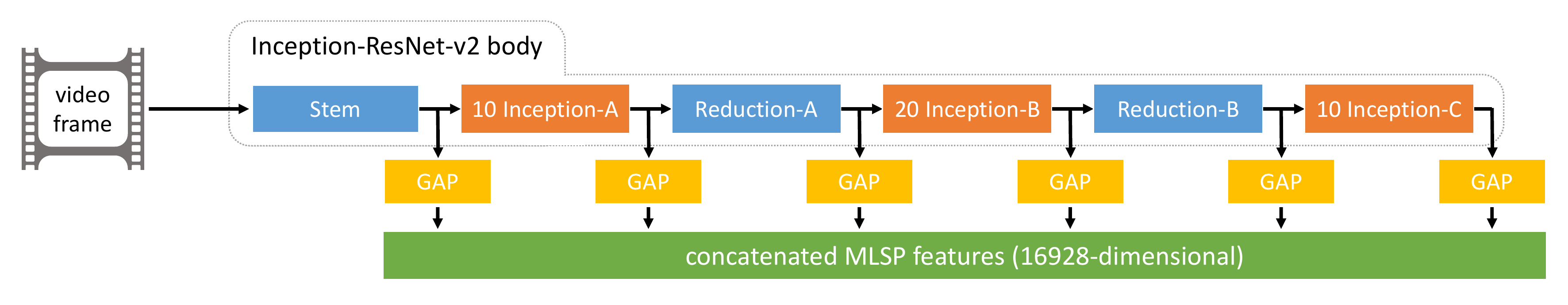} ~ 
		\caption{Extraction of multi-level spatially-pooled (MLSP) features from a video frame, using an InceptionResNet-v2 model pre-trained on ImageNet. The features encode quality-related information: earlier layers describe low-level image details, e.g. image sharpness or noise, and later layers function as object detectors or encode visual appearance information. Global Average Pooling (GAP) is applied to the activations resulting from the Stem, each Inception-module, as well as the Reduction-modules, and finally concatenated to form MLSP features. For more information regarding the individual blocks please refer to the original paper~\cite{szegedy2017inception}.}
		\label{fig:mlsp-feature-extraction}
	\end{figure*}

\begin{figure*}[h!]
		\centering
		\includegraphics[width=0.99\textwidth]{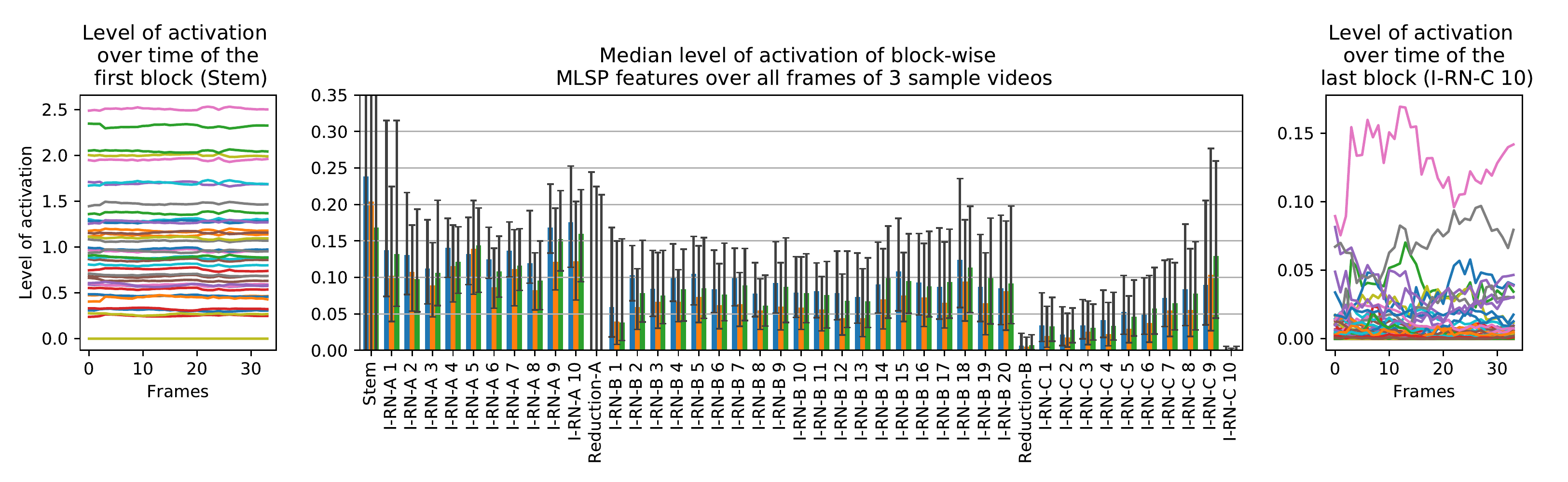} ~ 
		\caption{Visualization of the variation of activation levels of MLSP features over the course of KonVid-150k videos. In the center, the median level of activation for each of the 43 blocks from the Inception-ResNet-v2 network is displayed for 3 sample videos. The black whiskers indicate the 50\% confidence interval on the level of activation. For the first block (Stem), the whiskers extend to 0.7. The left and right plots show the activation of 1/8th of the first and last blocks' features over time.}
		\label{fig:mlsp-vis}
	\end{figure*}

	Instead of performing fine-tuning, we trained our models on features extracted from pre-trained DCNNs. The procedure is an expansion of what we described earlier for the comparison of content diversity, except we extracted features of all Inception modules of the network. The approach is inspired by \cite{hosu2019effective}, namely we extracted narrow multi-level spatially-pooled (MLSP) features, but for individual frames of videos, as shown in Fig.\ \ref{fig:mlsp-feature-extraction}. In principle, this general approach of extracting activations from individual layers of a network can be applied to any popular architecture. Related work has shown that this approach works with an Inception-ResNet-v2 network as a feature extractor in the IQA domain~\cite{hosu2020koniq,lin2020deepfl}. For the extraction process we, therefore, passed individual video frames to an InceptionResNet-v2 network, pre-trained on ImageNet \cite{szegedy2017inception}. We then performed global average pooling on the activation maps of all kernels in the stem of the network, as well as on each of the 40 Inception-ResNet modules and the two reduction modules. Concatenating the results yielded our MLSP feature vector consisting of average activation levels for 16,928 kernels of the InceptionResNet-v2 network. These MLSP feature vectors were extracted for all frames of all videos. Figure~\ref{fig:mlsp-vis} shows a visualization of parts of the MLSP feature vector for multiple consecutive frames.
	
	\subsection{Model Implementation Details}
	
\begin{figure*}[t!]
		\centering
		\includegraphics[width=0.8\textwidth]{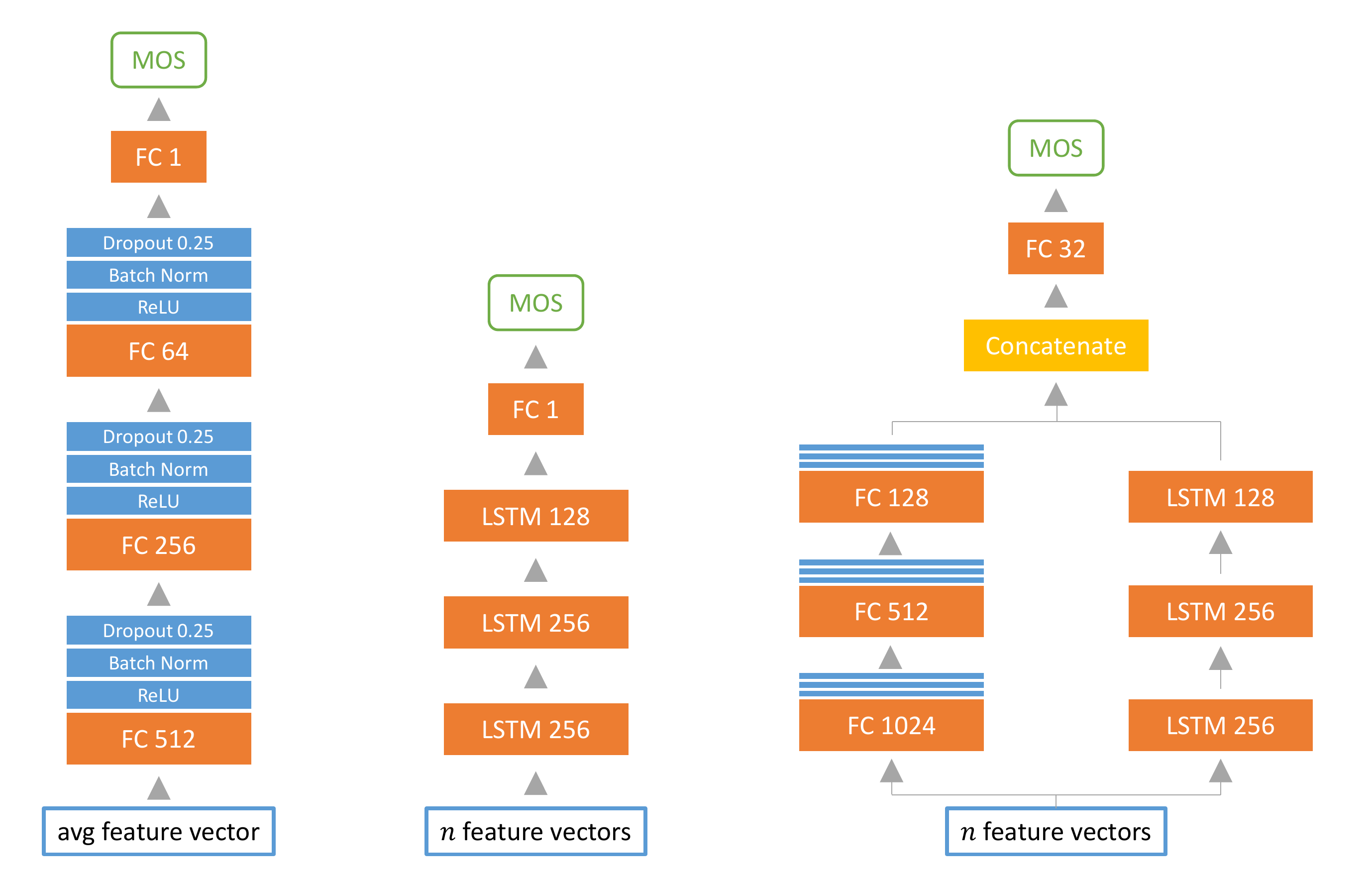} ~ ~
		\caption{Left: The MLSP-VQA-FF model, that relies on average frame MLSP features and a densely connected feed forward network. Middle: The MLSP-VQA-RN recurrent model, implementing a stacked long short-term memory network. Right: The hybrid MLSP-VQA-HYB dual channel model, that has a bigger variant of the FF network on the left and the recurrent part of the RN network on the right. Both channels output activations at each timestep and are merged along the feature dimension, before feeding into a small prediction head. Both the RN and HYB models take corresponding frame features at each time step as an input to the network.}
		\label{fig:proposednetworks}
	\end{figure*}

	Different learning-based regression models, such as Support Vector Regression (SVR) or Random Forest Regression (RFR), have been employed to predict subjective quality scores from frame features, with SVR yielding generally better results~\cite{korhonen2019hierarchical}. However, most existing works only extract a few dozen to a few hundred features. Since SVR is sub-optimal when applied to very large dimensional features like our MLSP feature, we instead train three small-capacity DNNs (Figure~\ref{fig:proposednetworks}): 
	\begin{itemize}
		\item MLSP-VQA-FF, a feed-forward DNN where the average feature vector is the input of three blocks of fully connected layers with ReLU activations, followed by batch normalization and dropout layers.
		\item MLSP-VQA-RN, a deep Long Short-Term Memory (LSTM) architecture, where each LSTM layer receives the feature vector or the hidden state of the lower LSTM layer as an input and outputs its hidden state. This stacking of layers allows for the simultaneous representation of input series at different time scales~\cite{hermans2013training}. The bottom LSTM layer can be understood as a selective memory of past feature vectors. In contrast, each additional LSTM layer represents a selective memory of past hidden states of the previous layer.
		\item MLSP-VQA-HYB, a two-channel hybrid of both the FF and RN variants. The temporal channel is a copy of the RN model's architecture, while the second channel is a mirror of the FF network scaled up to match the number of kernels in the temporal branch in the last layer. The outputs of the two channels are concatenated and a small 32 kernel fully connected layer feeds into the last prediction layer.
	\end{itemize}
	Our tests showed that employing dropout of any kind within the recurrent networks, such as input/output dropout or recurrent dropout, resulted in reduced performance. We therefore do not employ any dropout in these architectures.
	
	As mentioned before, this two-step strategy of feature extraction followed by training a regressor is much faster than transfer learning and fine-tuning an Inception-style network. It's difficult to fairly assess the difference, as a lot of factors play a role. For example, when fine-tuning an Inception-net, the speed at which the videos are read from the hard-drive can become a bottle-neck, if a very powerful GPU is performing the training procedure. Our proposed approach with an Inception-ResNet-v2 as a feature extraction network has a benefit for this scenario, since the input data for each frame is fixed at 16,928 floating point values. In contrast, if the GPU used to perform the training is not as powerful, it itself can become a bottle-neck of the system. In this case, our proposed approach has the alternative benefit that the small network size allows for much larger batches and quicker forward and backward passes.
	
	In order to quantify the difference, we compare different setups of transfer learning and  fine-tuning to our proposed two-step MLSP feature-based training procedure on a machine that reads from an NVMe connected SSD and trains the networks using Tensorflow 2.4.1 on an NVIDIA A100 with 40GB of VRAM. To simplify the setup, we are evaluating only the MLSP-VQA-FF model on the pre-extracted first frames of KonVid-150k-B. The transfer learning scenarios are all performed using an Inception-ResNet-v2 base-model with our FF model sitting on top for 40 epochs. However, we compare four slightly different scenarios:
	
	\begin{itemize}
		\item \textbf{Koncept}: The FF model takes the last layer of the base-model as an input, much like the Koncept model proposed in \cite{hosu2020koniq}. The weights of the base-model are not frozen, so the entire model is fine-tuned over the course of the training. We employ two training stages, one with a learning rate of $1\times10^{-3}$, and the second with a learning rate of $1\times10^{-5}$. 
		\item \textbf{IRNV2}: Instead of fine-tuning the entire model throughout both stages, we freeze the layers of the Inception-ResNet-v2 base-model for the first stage, so as to avoid the large update steps caused by the random initialisation of the header network to destroy the useful features in it. For the second stage we unfreeze the weights in all layers. 
		\item \textbf{IRNV2-MLSP}: As stated before, one downside of the above approaches lies in the circumstance that the header network relies only on the top level features as inputs. For the third comparison we concatenate the activation layers of all Inception-modules and feed that as an input to the header network. Here, we also freeze the base-model weights for the first stage, and unfreeze all weights for the second stage.
		\item \textbf{MLSP}: The final item in the comparison takes the MLSP features described above as an input. This means, the model is much smaller, as the base-model does not need to be loaded. However, the model can not leverage the spatial information about the activations to make it's prediction. No explicit weight freezing is performed in this scenario.
	\end{itemize}
	
\begin{figure}[t!]
		\centering
		\includegraphics[width=0.485\textwidth]{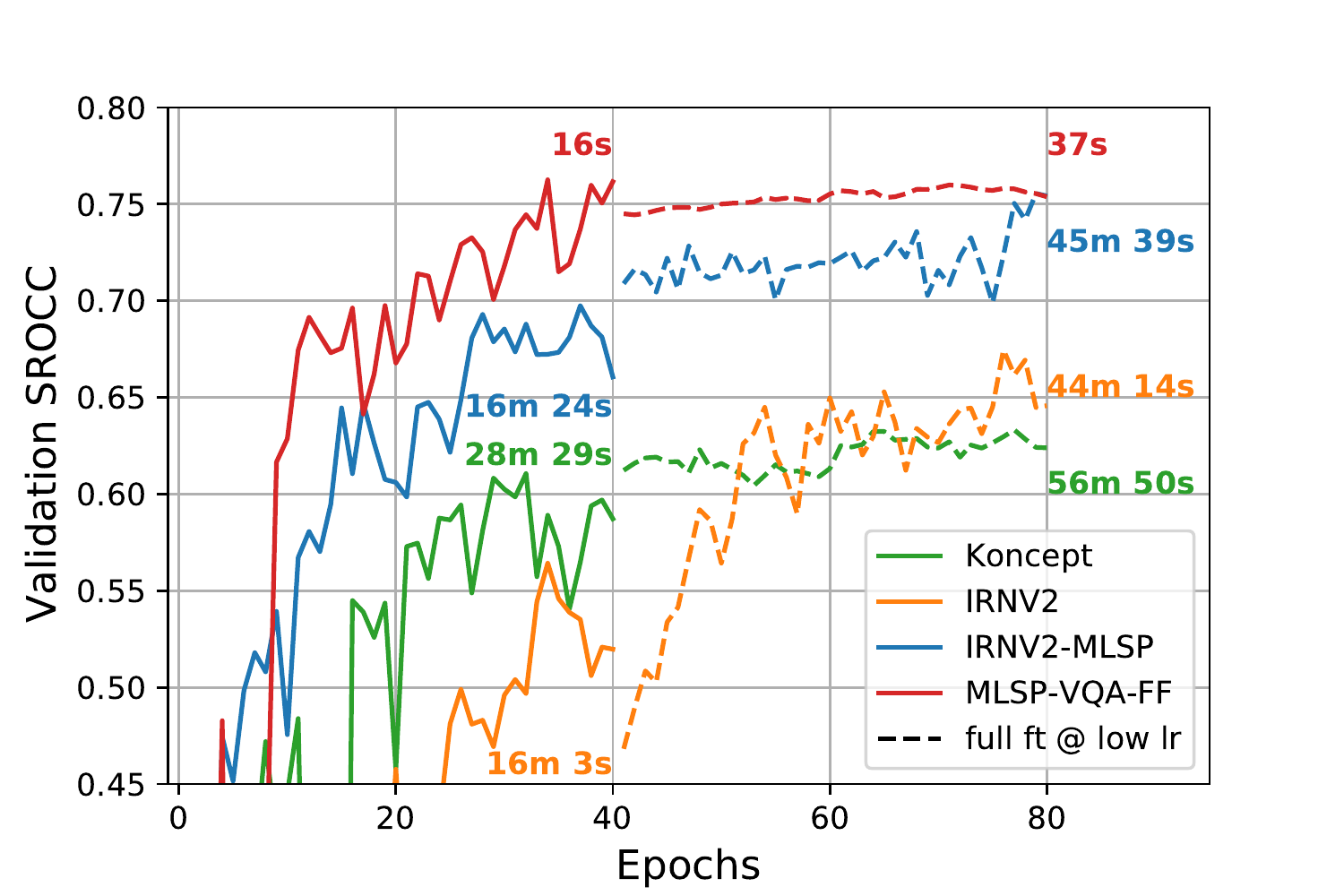} ~ 
		\caption{A visualization of the convergence of different transfer learning techniques along with information about the training times. The solid lines show the first training stage of 40 epochs, where the IRNV2 (orange) and IRNV2-MLSP (blue) architectures have their weights frozen. Koncept (greeN) and IRNV2 connect the last layer to the small header network, while IRNV2-MLSP concatenates all individual Inception-module outputs to feed into the head. Finally, MLSP-VQA-FF works off of extracted MLSP features, which for this scenario took 38 seconds.}\label{fig:db:mlspvsft}
	\end{figure}

	These different cases are compared in Figure~\ref{fig:db:mlspvsft}. The green graph, corresponding to the Koncept model, takes the longest to train in total and achieves the worst validation performance at the end of the 80 epochs. The reason for the slow training in the first stage is that none of the weights are frozen and the backpropagation step therefore takes additional time. Both the orange IRNV2 and blue IRNV2-MLSP models train faster by approximately 22\%, as the weights are frozen in the first stage. However, they differ in that the inclusion of all Inception-modules in the concatenation layer for the latter increases performance significantly. Finally, the red graph, representing the MLSP-VQA-FF model trained on extracted MLSP features achieves the best performance while beating the IRNV2-MLSP model in terms of speed by factor 74. Moreover, peak performance is achieved much earlier, as the second training stage is not required, raising the speed-up to factor 171. 
	
	However, feature extraction has to be performed once as well, which for the first frames of KonVid-150k-B took 38 seconds. Including this time in the comparison still renders the MLSP-VQA-FF model faster by factor 36, when considering both training stages. This factor is dependant on input resolutions, however with videos increasing in resolution the speed-up will only change in favor of the MLSP-based model, as its training speed will not change, while the training speed of the fine-tuning approach is inversely correlated with input resolution. This shows the power of using pre-extracted MLSP features. 
	
	Furthermore, we have observed the success of fine-tuning an Inception-style network in this manner is very sensitive to hyperparameters, while training the small FF network on MLSP features is fairly robust. 
	
	Table~\ref{tab:trainig-params} gives an overview of some hyperparameter settings used in the training of our MLSP-based models for the compared datasets. Mean square error (MSE) was used as a loss function for a duration of 250 epochs, stopping early if the validation loss did not improve in the most recent 25 epochs at an initial learning rate of $10^{-4}$. By default, the MLSP-VQA-FF model was trained with a learning rate of $10^{-2}$, and both the MLSP-VQA-RN and the MLSP-VQA-HYB models were trained with a learning rate of $10^{-4}$. 
	
	\begin{table}
		\caption{Training settings and parameters}
		\label{tab:trainig-params}
		\centering
		\resizebox{0.95\columnwidth}{!}{%
			\begin{tabular}{l|ccc|ccc}
				\toprule
				& \multicolumn{3}{c}{MLSP-VQA-FF}   & \multicolumn{3}{c}{MLSP-VQA-RN/-HYB} \\
				Type        & frames& batch size& lr        & frames & batch size& lr     \\
				\midrule
				KoNViD-1k     & all   & 128       & $10^{-2}$ & 180    & 128       & $10^{-4}$ \\
				LIVE-Qualcomm & all   & 8         & $10^{-3}$ & 150    & 8         & $10^{-4}$ \\
				CVD2014       & all   & 8         & $10^{-3}$ & 140    & 8         & $10^{-4}$ \\
				LIVE-VQC      & all   & 8         & $10^{-3}$ & 150    & 8         & $10^{-4}$ \\
				Proposed      & all   & 128       & $10^{-2}$ & 180    & 128       & $10^{-4}$ \\
				\bottomrule
			\end{tabular}%
		}
	\end{table}
	
	\section{Model Evaluation}
	\label{sec:modelevaluation}
	
	Our proposed NR-VQA approach of extracting features from a pre-trained classification network and training DNN architectures on them have been designed to predict video quality in-the-wild. We evaluate the potential of the MLSP features when used for training the shallow feed-forward and recurrent networks by measuring their performance on four widely used datasets (KoNViD-1k, LIVE-VQC, CVD2014, and LIVE-Qualcomm) and our newly established dataset KonVid-150k. We consider two basic scenarios, namely (1) intra-dataset, i.e. training and testing on the same dataset, and (2) inter-dataset, i.e., training (and validating) on our large dataset KonVid-150k and testing on another.
	
	There are two fundamental limitations in these datasets that affect the performance of our approach. The first one relates to the video content, in the form of domain shifts between ImageNet and the videos in the datasets. The other one is due to the different types of subjective video quality ratings (labels) in the datasets, that may affect the cross-testing performance.
	
	First, the features in the pre-trained network have been learnt from images in ImageNet. There are situations when the information in the MLSP features may not transfer well to video quality assessment: 
	\begin{itemize}
		\item Some artifacts are unique to video recordings; this is the case of temporal degradations such as camera shake, which does not apply to photos.
		\item Compression methods are different for videos in comparison to images. Thus, the individual frames may show encoding-specific artifacts that are not within the domain of artifacts present in ImageNet.
		\item In-the-wild videos have different types and magnitudes of degradations compared to photos. For example, motion blur degradations can be more prevalent and of a higher magnitude in videos compared to photos. This could affect how well MLSP features from networks pretrained on ImageNet transfer to VQA.
	\end{itemize}
	
	Secondly, concerning the subjective video quality ratings to be predicted when cross-testing, while there are similarities between the rating scales used in the subjective studies corresponding to each dataset, the ratings themselves may suffer from a presentation bias. For example, in the case of a dataset with highly similar scenes, but minuscule differences in degradation levels, as is the case for LIVE-Qualcomm and CVD2014, a human observer may become very sensitive to particular degradations. Conversely, video content becomes less critical for quality judgments. The attention of the human observer is diverted to parts in the video he might otherwise not have looked at, had he not seen the same or a very similar scene many times before. Whether the resulting subjective judgments can be regarded as fair quality values is arguable. A human observer would rarely watch a scene multiple times before rating the quality. This bias of subjective opinions may greatly influence how the quality predictions trained in one setting generalize to others. Similarly, quality scores obtained in a lab environment will be much more sensitive to differences in technical quality than a worker in a crowdsourcing experiment might be able to pick up. Therefore, it may be challenging to generalize from one experimental setup to another. While consumption of ecologically valid video content happens in a variety of environments and on a multitude of devices, it is arguable whether one experimental setup is superior.
	
	\subsection{Model Performance Comparisons}
	
	We first evaluate the performance of the proposed model on four existing video datasets. KoNViD-1k and LIVE-VQC both pose the unique challenge that they are in-the-wild video datasets, containing authentic distortions that are common to videos hosted on Flickr. LIVE-Qualcomm contains self-recorded scenes of different mobile phone cameras that were aimed at inducing common distortions. CVD2014 differs from the previous two, in that it is a dataset with artificially introduced acquisition-time distortions. It also contains only five unique scenes depicting people. Finally, LIVE-VQC was a collaborative effort of friends and family of the LIVE research group that were asked to submit video files of a variety of contents to capture diversity in capturing equipment and distortions.
	
	We are comparing our proposed DNN models against published results for other methods that have been thoroughly evaluated on these datasets using SVR and RFR. Detailed information regarding the experimental evaluation and results of the classical methods can be found in \cite{korhonen2019hierarchical}. 
	
	We adopt a similar testing protocol by training 100 different random splits with 60\% of the data used for training, 20\% used for validation, and 20\% for testing in each split. Table~\ref{tab:datasetresults} summarizes the SRCC w.r.t.\ the ground-truth for the predictions of the classical methods (taken from~\cite{korhonen2019hierarchical}) alongside our DNN-based approach. It is to be noted that the random splits we used are different from the ones used to evaluate the classical methods in \cite{korhonen2019hierarchical}. For brevity, we are only reporting the results for classical methods obtained using SVR, although four individual results are slightly improved using RFR. 
	
	The FF network outperforms the existing works on KoNViD-1k, improving state-of-the-art SRCC from 0.80 to 0.82, while the RN and HYB models remain competitive with an SRCC of 0.78 and 0.79, respectively. This shows that the proposed approaches are performing close to state-of-the-art on authentic videos with some encoding degradations. Since the feature extraction network is trained on images with natural image distortions, some of the extracted features are likely indicative of these distortions, which are not unlike the video encoding artifacts introduced by Flickr.
	
\begin{figure}[b!]
		\centering
		\includegraphics[width=0.45\textwidth]{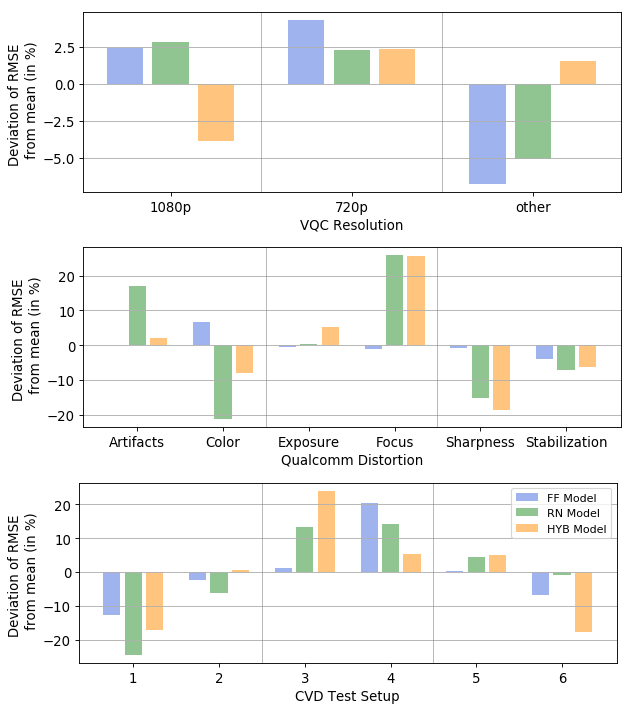} ~
		\caption{Percent deviation of the mean RMSE of the proposed models on each of the six degradation types present in LIVE-Qualcomm (top), each of the six test scenarios in CVD2014 (middle), and the different resolutions in LIVE-VQC (bottom).}\label{fig:res:rmse-deviation}
	\end{figure}

	Existing methods had not been evaluated exhaustively on LIVE-VQC at the time of writing. Our recurrent networks achieve 0.70 (RN) and 0.69 (HYB) SRCC, while the FF model performs at 0.72 SRCC, rendering it competitive with state-of-the-art for the dataset\footnote{\label{note1}Recently, a new publication on arXiv disusses a new approach called RAPIQUE that achieves an SROCC of 0.76 on LIVE-VQC\ref{tu2021rapique}. However, this work has not yet been peer reviewed.}. One of the difficulties inherent to VQC with respect to our models is the circumstance, that it is comprised of videos of various resolutions and aspect ratios. An evaluation of the performance of the models with respect to the video resolutions can be found in the top part of Figure~\ref{fig:res:rmse-deviation}. Since 1080p, 720p, and 404p in portrait orientation are the predominant resolutions with 110, 316, and 119 videos, respectively, we grouped the other resolutions into the \textit{other} category. We can see that both the FF and RN models perform worse on the 1080p and 720p videos, whereas the HYB model performs better on the higher resolution videos.
	
	In the case of LIVE-Qualcomm our best performance of 0.75 SRCC of the hybrid model is surpassed only by TLVQM with 0.78. Since the dataset is comprised of videos containing six different distortion types, we also evaluated the performance of the models according to each degradation, as depicted in the middle plot of Figure~\ref{fig:res:rmse-deviation}. Here, we show the deviation of the RMSE of each model for each distortion type from the average performance in percent. Little deviation between all three models is observed for both Exposure and Stabilization type distortions. However, for Artifacts and Color the RN model deviates from the other two drastically, performing worse on the former and better on the latter. Videos in the focus degradation class show auto-focus related distortions where parts of the video are intermittently blurry or sharp over time and are overall the biggest challenge for our recurrent models, that both perform over 20\% worse on them than average. Finally, the Sharpness distortion is best predicted by the recurrent networks, with the hybrid model outperforming the pure LSTM network.
	
	On CVD2014, our proposed models with SRCCs of 0.77, 0.75, and 0.79 for the FF, RN and HYB models, respectively, are outperformed by both FRIQUEE and TLVQM at 0.82 and 0.83 SRCC. CVD2014 is a dataset of videos of two different resolutions, with artificially introduced capturing distortions and only five unique scenes of humans and human faces. The magnitude of the artifacts is at a level that is not commonly seen in videos in-the-wild, and the types of defects are also not within the domain of distortions present in ImageNet. Therefore, this is the most challenging dataset for our approach and, consequently, the relative performance of our approach is worse. CVD2014 is split into six subsets with partially overlapping scenes but distinct capturing cameras. The bottom part of Figure~\ref{fig:res:rmse-deviation} shows the relative deviation of the RMSE from the mean performance for each of these test setups. The first two setups include videos at 640$\times$480 pixels resolution, which are generally rated with a lower MOS than videos in the other test setups, which could both be an important factor in our models' increased performance here. Although all setups include scenes 2 and 3, scene 1 is only included in test setups 1 and 2, scene 4 is only included in test setups 3 and 4, and scene 5 is solely included in test setups 5 and 6. Since the features we use are tuned to identify content, as we showed in Section~\ref{sec:db:eval}, inclusion or exclusion of particular scenes can have an impact on the performance of our method. Moreover, since each test setup contains videos taken from different cameras than the rest, it is possible that the in-capture distortions caused by particular cameras in any individual test setup may be closer to the types of distortions present in ImageNet.
	
	\begin{table*}[h!]
		\caption{Results of different NR-VQA metrics on different authentic VQA datasets}
		\label{tab:datasetresults}
		\centering
		\begin{threeparttable}
			\begin{tabular}{clccccc}
				&                       & \multicolumn{2}{c}{in-the-wild}                         & & \multicolumn{2}{c}{synthetic}                   \\
				\cmidrule{3-4}\cmidrule{6-7}
				&                       & KoNViD-1k                 & LIVE-VQC                    & & LIVE-Qualcomm              & CVD2014                  \\
				& Name                  & SRCC ($\pm\sigma$)        & SRCC ($\pm\sigma$)          & & SRCC ($\pm\sigma$)         & SRCC ($\pm\sigma$)  \\
				\cmidrule{1-7}
				\parbox[t]{2mm}{\multirow{7}{*}{\rotatebox[origin=c]{90}{SVR}}}
				& NIQE (1 fps)          & 0.34 ($\pm$0.05)         & 0.56 ($\pm$--.--{}--)      & & 0.46 ($\pm$0.13)          & 0.58 ($\pm$0.10)        \\
				& BRISQUE (1 fps)       & 0.56 ($\pm$0.05)\tnote{1}& 0.61 ($\pm$--.--{}--)      & & 0.55 ($\pm$0.10)          & 0.63 ($\pm$0.10)\tnote{1} \\
				& CORNIA (1 fps)        & 0.51 ($\pm$0.04)         & --.--{}-- ($\pm$--.--{}--) & & 0.56 ($\pm$0.09)          & 0.68 ($\pm$0.09)       \\
				& V-BLIINDS             & 0.65 ($\pm$0.04)\tnote{1}& 0.72 ($\pm$--.--{}--)      & & 0.60 ($\pm$0.10)          & 0.70 ($\pm$0.09)\tnote{1} \\ 
				& HIGRADE (1 fps)       & 0.73 ($\pm$0.03)         & --.--{}-- ($\pm$--.--{}--) & & 0.68 ($\pm$0.08)          & 0.74 ($\pm$0.06)       \\
				& FRIQUEE (1 fps)       & 0.74 ($\pm$0.03)         & --.--{}-- ($\pm$--.--{}--) & & 0.74 ($\pm$0.07)          & 0.82 ($\pm$0.05)       \\
				& TLVQM                 & 0.78 ($\pm$0.02)         & --.--{}-- ($\pm$--.--{}--) & & \textbf{0.78 ($\pm$0.07)} & \textbf{0.83 ($\pm$0.04)} \\
				\cmidrule{1-7}
				\parbox[t]{2mm}{\multirow{4}{*}{\rotatebox[origin=c]{90}{DNN}}}
				& 3D-CNN + LSTM\tnote{2}& 0.80 ($\pm$--.--{}--)    & --.--{}-- ($\pm$--.--{}--) & & 0.69 ($\pm$--.--{}--)     &  --.--{}-- ($\pm$--.--{}--) \\
				& MLSP-VQA-FF           & \textbf{0.82 ($\pm$0.02)}& \textbf{0.72 ($\pm$0.06)}  & & 0.71 ($\pm$0.08)          & 0.77 ($\pm$0.06)      \\
				& MLSP-VQA-RN           & 0.78 ($\pm$0.02)         & 0.70 ($\pm$0.06)           & & 0.72 ($\pm$0.07)          & 0.75 ($\pm$0.06)      \\
				& MLSP-VQA-HYB          & 0.79 ($\pm$0.02)         & 0.69 ($\pm$0.07)           & & 0.75 ($\pm$0.04)          & 0.79 ($\pm$0.05)      \\
				\bottomrule
			\end{tabular}
			\begin{tablenotes}
				\item[1] Performance improves when using random forest regression. 
				\item[2] The authors did not supply any standard deviations for the performance measures, and did not evaluate the method on CVD2014.
			\end{tablenotes}
		\end{threeparttable}
	\end{table*}
	
	We now consider the performance evaluation when training and testing on our new dataset, KonVid-150k-B of 1,596 videos, each with at least 89 ratings comprising the quality score. We separate these tests from the previous ones because, in this case, we have the option to train the networks on the additional 150k videos in KonVid-150k-A that stem from the same domain. From the previous experiments, it is evident that TLVQM is the best performing classical metric on the similar domain, given by KoNViD-1k, by a large margin. Therefore, we compare our MLSP-VQA models only against TLVQM and the standard V-BLIINDS.
	
	Table~\ref{tab:k1600} summarizes the performance results. Compared to the performance on KoNViD-1k, V-BLIINDS (row 1) improves slightly, while TLVQM (row 2) performs significantly worse. Since the main difference between KoNViD-1k and this dataset is the reduced re-encoding degradations, it appears as though the classical methods over-emphasize their prediction on these artifacts. The third through fifth row list the performance of our models, which outperform both classical methods, beating TLVQM's 0.71 SRCC with 0.83 (FF), 0.78 (RN) and 0.75 (HYB) when trained and tested on the B variant exclusively. 
	
	Finally, the last three rows show the results from training on the large dataset, KonVid-150k-A, with 150k videos. For these last three evaluations a random subset of 50\% of KonVid-150k-B was used for validation during training. The remaining part of KonVid-150k-B was used for testing. We note an additional substantial performance increase for our networks. The FF model's performance increases from 0.81 SRCC to 0.83, while the RN model improves from 0.78 SRCC to 0.81. The largest performance gain can be observed for the HYB network, as it improves from 0.75 SRCC to 0.81 SRCC as well. This demonstrates, for the first time, the enormous potential gains that can be achieved by vast training datasets for VQA. Although KonVid-150k-A only has MOS scores comprised of five individual votes, by training on them and validating on the target dataset we drastically improve performance. It is to be noted as well that the test sets in this scenario are larger than when training and testing solely on KonVid-150k-B. This renders the test performance to be even more representative. However, the change in variance of the resulting correlation coefficients can not directly be attributed to the increase in training dataset size. The difference likely arises from the fact that the models trained using KonVid-150k-A have the same training data, and are therefore more likely to learn similar features. Nonetheless, this effect should be investigated further.
	
	\begin{table*}
		\caption{Results of NR-VQA metrics on KonVid-150k-B. The bottom three rows describe the performance when training on the entirety of KonVid-150k-A, using half of KonVid-150k-B as a validation set, and the other.}
		\label{tab:k1600}
		\centering
		\begin{tabular}{cllll}
			\toprule
			& Name                    & PLCC ($\pm\sigma$)  & SRCC ($\pm\sigma$)  & RMSE ($\pm\sigma$)   \\
			\midrule
			\parbox[t]{2mm}{\multirow{2}{*}{\rotatebox[origin=c]{90}{SVR}}}
			& V-BLIINDS (SVR)         & 0.68 ($\pm$0.04) & 0.68 ($\pm$0.04) & 0.27 ($\pm$0.02) \\ 
			& TLVQM (SVR)             & 0.68 ($\pm$0.12) & 0.71 ($\pm$0.04) & 0.26 ($\pm$0.04) \\
			\midrule
			\parbox[t]{2mm}{\multirow{6}{*}{\rotatebox[origin=c]{90}{DNN}}}
			& MLSP-VQA-FF             & 0.83 ($\pm$0.02) & 0.81 ($\pm$0.02) & 0.26 ($\pm$0.01) \\
			& MLSP-VQA-RN             & 0.80 ($\pm$0.02) & 0.78 ($\pm$0.02) & 0.29 ($\pm$0.01) \\
			& MLSP-VQA-HYB            & 0.76 ($\pm$0.04) & 0.75 ($\pm$0.04) & 0.32 ($\pm$0.03) \\
			\cline{2-5}
			& MLSP-VQA-FF (Full)      & 0.86 ($\pm$0.01) & 0.83 ($\pm$0.01) & 0.19 ($\pm$0.01) \\
			& MLSP-VQA-RN (Full)      & 0.83 ($\pm$0.01) & 0.81 ($\pm$0.01) & 0.21 ($\pm$0.01) \\
			& MLSP-VQA-HYB (Full)     & 0.83 ($\pm$0.01) & 0.81 ($\pm$0.01) & 0.21 ($\pm$0.01) \\
			\bottomrule
		\end{tabular}
	\end{table*}

	\subsection{Inter-Dataset Performance}
	
	Considering the diversity in content and distortions in KonVid-150k we highlight the power of KonVid-150k in combination with our MLSP-VQA models in inter-dataset testing scenarios. At the time of writing, LIVE-VQC has not been considered in any performance evaluations across datasets. The previously best reported cross-test performances between the other three legacy datasets are three different combinations of NR-VQA methods and training datasets\footnote{These results are taken from \cite{korhonen2018learning}.}. Specifically, TLVQM trained on CVD2014 performs best on KoNViD-1k cross-testing with 0.54 SRCC. V-BLIINDS trained on KoNViD-1k is the best combination for cross-testing on LIVE-Qualcomm with 0.49 SRCC. Finally, FRIQUEE trained on KoNViD-1k performs best when cross-testing on CVD2014 with 0.62 SRCC. It is apparent from these results that no single NR-VQA and dataset combination generally outperforms in inter-dataset testing scenarios. 
	
	\begin{table*}
		\caption{Inter-dataset test performance of our three models averaged over 10 splits trained on the entirety of KonVid-150k-A. The different splits only affect the validation and test sets, as all videos of KonVid-150k-A are used for training.}
		\label{tab:cross-dataset-tests}
		\centering
		\begin{tabular}{lcccccc}
			& \multicolumn{2}{c}{in-the-wild}                & & \multicolumn{2}{c}{synthetic}                   \\
			\cmidrule{2-3}\cmidrule{5-6}
			&  KoNViD-1k                 & LIVE-VQC                   & & LIVE-Qualcomm              & CVD2014            \\
			& SRCC ($\pm\sigma$)            & SRCC ($\pm\sigma$)            & & SRCC ($\pm\sigma$)            & SRCC ($\pm\sigma$)   \\
			\cmidrule{1-3}\cmidrule{5-6}
			Intra-dataset best &  0.82 ($\pm$0.02)      & 0.72 ($\pm$0.06)           & & 0.78 ($\pm$0.07)           & 0.83 ($\pm$0.04) \\
			\cmidrule{1-3}\cmidrule{5-6}
			Prev. inter-dataset best\cite{korhonen2018learning}    &  0.54 ($\pm$--.--{}--)     & --.--{}-- ($\pm$--.--{}--) & & 0.49($\pm$--.--{}--)       & \textbf{0.62 ($\pm$--.--{}--)} \\
			MLSP-VQA-FF    &  \textbf{0.83 ($\pm$0.01)} & \textbf{0.75 ($\pm$0.01)}  & & \textbf{0.64 ($\pm$0.01)}  & 0.55 ($\pm$0.02)  \\
			MLSP-VQA-RN    &  0.80 ($\pm$0.01)          & 0.71 ($\pm$0.01)           & & 0.61 ($\pm$0.03)           & 0.52 ($\pm$0.02)   \\
			MLSP-VQA-HYB   &  0.79 ($\pm$0.01)          & 0.71 ($\pm$0.01)           & & 0.62 ($\pm$0.03)           & 0.52 ($\pm$0.02) \\
			\bottomrule
		\end{tabular}
	\end{table*}
	
	We evaluate the performance of our models when cross-testing on other datasets, trained on KonVid-150k-A and validated and tested on each 50\% of KonVid-150k-B. The average SRCC performances of 10 models are reported in Table~\ref{tab:cross-dataset-tests}. For ease of comparison we also include the best within-dataset performance in the first row, as well as the previous best cross-dataset test performances as taken from \cite{korhonen2018learning} in the second row of the table. Although the performances between our different models do not vary much, the results reveal some interesting findings.
	
	\begin{itemize}
		\item The cross-dataset test performance of the FF model on KoNViD-1k of 0.83 SRCC is higher than all other within-dataset test performances and especially any cross-test setups. This again underlines the potential power of data, even if it is annotated with lower precision. Although KonVid-150k does not have the Flickr video encoding artifacts present, it can predict the distorted videos of KoNViD-1k better than training on videos taken from the same dataset. 
		\item Our models trained on KonVid-150k and cross-tested on LIVE-VQC achieve state-of-the art performance and even surpass the best within-dataset performance in the case of the FF model with 0.75 SRCC$^{\ref{note1}}$
		\item On LIVE-Qualcomm the cross-dataset test performances of all our models are slightly better than V-BLIINDS (0.60), when it is trained and tested on LIVE-Qualcomm. Since V-BLIINDS has been the de facto baseline method, this is a remarkable result. Additionally, for a cross-dataset test our proposed KonVid-150k dataset shows the best generalization to LIVE-Qualcomm, improving the previous best 0.49 SRCC to 0.64.
		\item Next, our models struggle with CVD2014, as none of them beat even the most dated classical models trained and tested on CVD2014 itself. This may be in part due to the nature of the degradations induced in the creation of the dataset, which are not native to the videos present in KonVid-150k. Moreover, the domain shift between KonVid-150k and CVD2014 seems to be larger than to the other datasets, as the previous best cross-dataset performance is also not achieved.
	\end{itemize}
	
	The cross-test performance drops notably when testing on synthetic video datasets. This has already been observed in the IQA domain~\cite{lin2020deepfl}, where training and testing on the same domain resulted in much higher performance than when the source and target domains were different. The types of distortions in individual frames of videos from two different domains result in different characteristics of the activations of Inception-net features, resulting in reduced performance.
	
	\subsection{Evaluation of Training Schemes}
	\label{sec:modeleval:trainingschemes}
	
	As described in Section~\ref{sec:rw:vqa-datasets}, the choice of the number of ratings per video is a distinguishing, yet so far unexplored factor in the design of VQA datasets in the context of optimizing model training performance. In order to study the effect of varying the number of ratings per video, we trained a large set of corresponding models in two experiments. In the first one, we increased the number of ratings to reduce the level of noise in the training set. In the second one, we additionally introduced the natural constraint of a vote budget, limiting the total number of ratings to a constant.
	
	It is common to use an equal number of votes for each stimulus so that the MOS of the training, validation, and test sets have the same reliability, respectively, the same level of noise. Deep learning is known to be robust to label noise~\cite{rolnick2017deep}, however, this has been only studied when the same amount of noise is present for all items in all parts of the dataset (train/test/validation). Thus, the first question we investigate is:
	\begin{itemize} \item \textit{What impact do different noise levels in the training and validation sets have on test set prediction performance?}\end{itemize}
	More precisely, we are interested to know the change in prediction performance when fewer votes are used for training and validating deep learning models, compared to the number of votes used for test items. 
	
	In order to answer this question, we randomly sampled $v = 1, 2, 4, 7, 14, 26,$ and $50$ votes five times for each video within KonVid-150k-B and computed the corresponding MOS values (7$\times$5 MOS per video). We then trained our MLSP-VQA-FF model by varying both training set, and validation set MOS vote counts while keeping the test set MOS vote count at 50. For each pair of training and validation MOS, we considered twenty random splits with 60\% of the data for training, 20\% for validation, and 20\% for testing, with the above mentioned five versions of the MOS each. Therefore, we trained $5 \times 20 \times 7 \times 7 = 4900$ models in total. 
	
\begin{figure}[t!]
		\centering
		\includegraphics[width=0.49\textwidth]{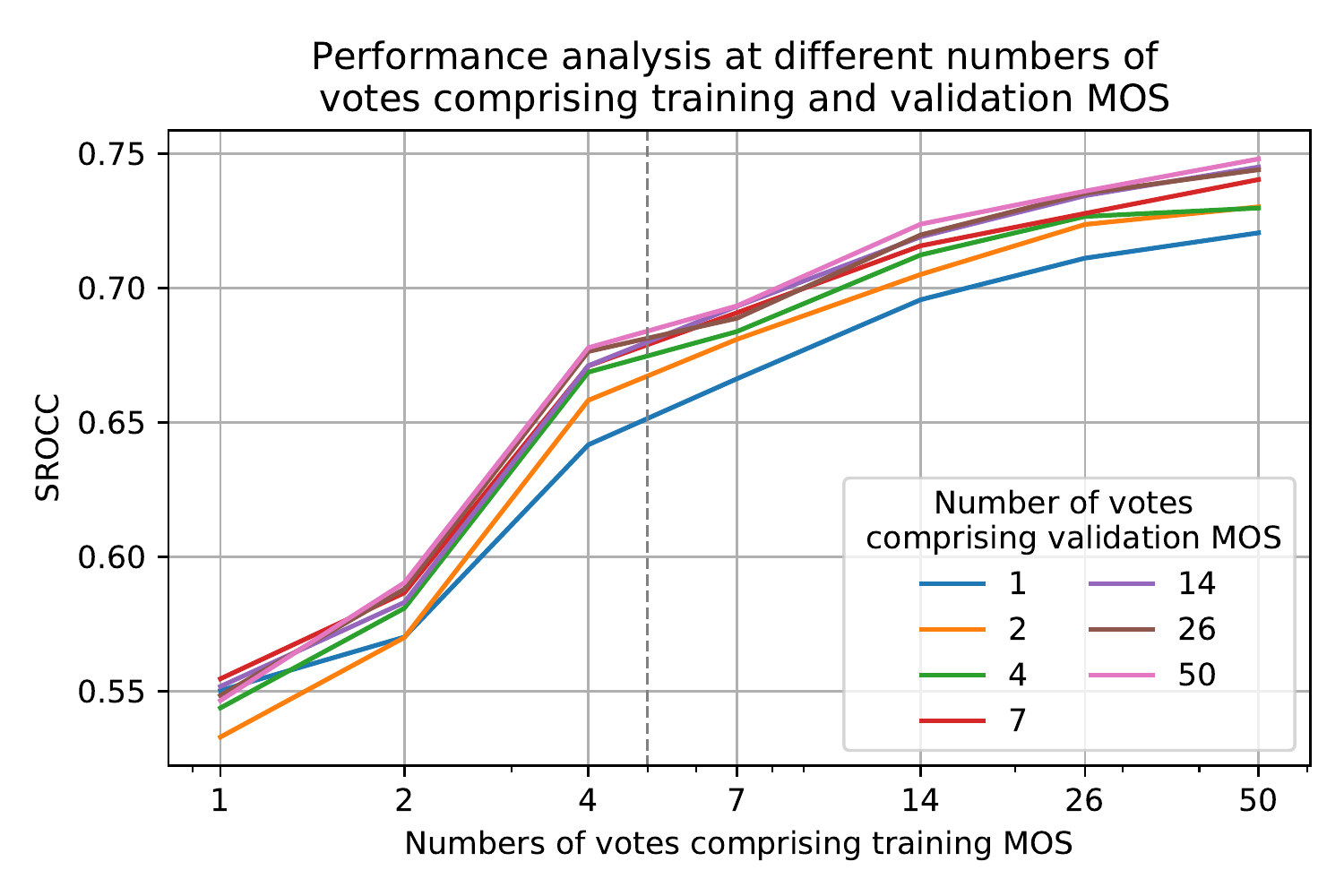} 
		\caption{Noise robustness of MLSP feature based method.}\label{fig:MLSPnoiserobustness}
	\end{figure}

	The graph in Figure \ref{fig:MLSPnoiserobustness} depicts the mean SRCC between the models' predictions and the ground truth MOS of the test sets. Each line in this graph represents a different number of votes comprising the validation MOS, whereas the x-axis indicates the number of votes comprising the training MOS. Note that the x-axis is scaled logarithmically for better visualization. There are three key observations concerning the prediction performance:
	
	\begin{itemize}
		\item The prediction performance improves as the number of votes comprising the training MOS increases, regardless of the number of votes used for validation.
		\item The performance improvements scale approximately logarithmically with the number of votes comprising the training MOS.
		\item The test set performance varies less due to changes in the number of votes used for validation than it does due to the number of votes for items in the training set.
	\end{itemize}
	
	The fact that performance improves with lower training label noise is not surprising. Nonetheless, the gentler slope for the performance curves beyond four votes comprising the training MOS is an indicator that the common policy to gather 25 votes for all stimuli in a dataset may be sub-optimal, due to diminishing returns. In fact, at approximately five votes (1/10th of the analysed budget) the model achieves roughly 92\% of the peak performance, suggesting it to be a good trade-off between precision and cost.
	
	The comparison between data splits in this experiment is not balanced, because the data points in the graphs of Figure~\ref{fig:MLSPnoiserobustness} correspond to different vote budgets, ranging from 1 rating per video in one instance on the left up to 50 per video on the right. The annotation of datasets in the lab and also in the crowd usually is constrained by a budget in terms of total hours of testing or overall cost of crowdsourcing. This translates to a maximum number of votes that can be attained for a given dataset. Therefore, the second question we investigate is:
	\begin{itemize} \item \textit{Given a fixed vote budget, how does the allocation of votes on the training set affect test performance?}
	\end{itemize}
	In other words, is it better to collect more votes for fewer stimuli, or less votes for more videos?
	
	In order to answer this question, we first divided KonVid-150k-B into five disjoint test sets (each with 20\% of all videos) and sampled the same number of videos from the remaining set of KonVid-150k-B for validation. We then considered three levels of precision at 100, 5, and 1 votes comprising the MOS of videos used in training, as well as six vote budgets of 100,000, 25,000, 10,000, 2,500, and 1,000 votes. We built the training sets accordingly, sampling from the remaining videos in KonVid-150k-B first, and then adding in videos from KonVid-150k-A, if needed, such that the smaller sets are proper subsets of the larger variants. For the vote budget of 100,000 votes we consequently created three training sets of 1,000, 20,000, and 100,000 videos at training MOS precision levels of 100, 5 and 1 vote(s), respectively. It is to be noted that the overlap between the different samples of the same sets increases as the set size increases, as the whole KonVid-150k-B set is only comprised of $\approx$150,000 videos, which in turn has an effect on the standard deviation of the predictions.
	
	We trained both MLSP-VQA-FF and MLSP-VQA-RN on the five different splits for all three vote budget distributions and reported the results in Table~\ref{tab:mlsp-votebudget}. We give the average SRCC, PLCC, and RMSE between the models' predicted scores and the MOS computed by using all available votes. There are are few key takeaways from these results:
	
	\begin{itemize}
		\item As one would suspect, the performance drops as the total vote budget decreases. 
		\item Surprisingly, however, the performance appears to be stable across the different distribution strategies for budgets of more than 1,000 votes.
		\item For smaller vote budgets a middle ground choice between MOS precision and numbers of videos seems to be favorable, as indicated by the 5 vote MOS distribution strategy outperforming the more and less precise extreme strategies. This suggests that for very small vote budgets in particular the focus should be on fewer than the commonly suggested 30 rating MOS recommendations that are found in literature.
	\end{itemize}
	
	
	\begin{table}
		\caption{Performance of our FF model at a fixed vote budget of 100,000, 25,000, 10,000, 2,500, and 1,000 votes.}
		\label{tab:mlsp-votebudget}
		\centering
		\begin{tabular}{cccc}
			\toprule
			Set & PLCC & SRCC & RMSE \\
			\midrule
			\addstackgap{\alignCenterstack{\phantom{00}1000&@100}} & 0.76 ($\pm$0.03) & 0.73 ($\pm$0.04) & 0.24 ($\pm$0.01) \\
			\addstackgap{\alignCenterstack{\phantom{0}20000&@5\phantom{00}}}  & 0.76 ($\pm$0.02) & 0.74 ($\pm$0.03) & 0.24 ($\pm$0.01) \\
			\addstackgap{\alignCenterstack{100000&@1\phantom{00}}} & 0.77 ($\pm$0.02) & 0.74 ($\pm$0.03) & 0.24 ($\pm$0.01) \\
			\midrule
			\addstackgap{\alignCenterstack{\phantom{000}250&@100}} & 0.75 ($\pm$0.01) & 0.70 ($\pm$0.01) & 0.26 ($\pm$0.01) \\
			\addstackgap{\alignCenterstack{\phantom{00}5000&@5\phantom{00}}}  & 0.77 ($\pm$0.02) & 0.72 ($\pm$0.02) & 0.25 ($\pm$0.01) \\
			\addstackgap{\alignCenterstack{\phantom{0}25000&@1\phantom{00}}} & 0.76 ($\pm$0.02) & 0.72 ($\pm$0.02) & 0.25 ($\pm$0.01) \\
			\midrule
			\addstackgap{\alignCenterstack{\phantom{000}100&@100}} & 0.68 ($\pm$0.03) & 0.62 ($\pm$0.05) & 0.28 ($\pm$0.01) \\
			\addstackgap{\alignCenterstack{\phantom{00}2000&@5\phantom{00}}}  & 0.68 ($\pm$0.02) & 0.64 ($\pm$0.03) & 0.28 ($\pm$0.02) \\
			\addstackgap{\alignCenterstack{\phantom{0}10000&@1\phantom{00}}} & 0.69 ($\pm$0.06) & 0.66 ($\pm$0.05) & 0.28 ($\pm$0.01) \\
			\midrule
			\addstackgap{\alignCenterstack{\phantom{0000}25&@100}} & 0.56 ($\pm$0.08) & 0.51 ($\pm$0.07) & 0.32 ($\pm$0.02) \\
			\addstackgap{\alignCenterstack{\phantom{000}500&@5\phantom{00}}}  & 0.59 ($\pm$0.04) & 0.54 ($\pm$0.07) & 0.34 ($\pm$0.02) \\
			\addstackgap{\alignCenterstack{\phantom{00}2500&@1\phantom{00}}} & 0.57 ($\pm$0.04) & 0.52 ($\pm$0.05) & 0.36 ($\pm$0.04) \\
			\midrule
			\addstackgap{\alignCenterstack{\phantom{0000}10&@100}} & 0.46 ($\pm$0.07) & 0.41 ($\pm$0.09) & 0.34 ($\pm$0.02) \\
			\addstackgap{\alignCenterstack{\phantom{000}200&@5\phantom{00}}}  & 0.55 ($\pm$0.05) & 0.50 ($\pm$0.07) & 0.34 ($\pm$0.02) \\
			\addstackgap{\alignCenterstack{\phantom{00}1000&@1\phantom{00}}} & 0.46 ($\pm$0.12) & 0.44 ($\pm$0.10) & 0.45 ($\pm$0.05) \\
			\bottomrule
		\end{tabular}
	\end{table}

	\section{Conclusions}
	
	We introduced a large-scale in-the-wild dataset KonVid-150k for video quality assessment (VQA), as well as three novel state-of-the-art no-reference VQA methods for videos in-the-wild. Our learning approach (MLSP-VQA) outperforms the best existing VQA methods trained end-to-end on several datasets, and is substantially faster to train without sacrificing any predictive power. The large size of the database and efficiency of the learning approach have enabled us to study the effect of different levels of label-noise and how the vote budget (total number of collected scores from users) affects model performance. We were able to study the effect of different vote budget distribution strategies, meaning that the number of annotated videos was adjusted according to the desired MOS precision. Under a fixed budget, we found that in most cases the number of votes allocated to each video is not important for the final model performance when using our MLSP-VQA approach and other feature-based approaches.
	
	KonVid-150k takes a novel approach to VQA, going far beyond the usual in the VQA community. The database is two orders of magnitude larger than previous datasets, and it is more authentic both in terms of variety of content types and distortions, but also due to the compression settings of the videos. We retrieved the original video files uploaded by users from Flickr, without the default re-encoding that is generally applied by any video sharing platform to reduce playback bandwidth costs. We encoded the raw video files ourselves at a high enough quality to ensure the right balance between quality and size constraints for crowdsourcing.
	
	The main novelty of the proposed MLSP-VQA-HYB method is the two-channel architecture. By global average pooling the activation maps of all kernels in the Inception modules of an InceptionResNet-v2 network trained on ImageNet, we extract a wide variety of features, ranging from detections of oriented edges to more abstract ones related to object category. These features are input to the partially recurrent DNN architectures, which on the one hand makes use of the temporal sequence of the frame features, while on the other also considering the individual frame features as well. 
	
	We have trained and validated the proposed methods on the four most relevant VQA datasets, improving state-of-the-art performance on KoNViD-1k and LIVE-VQC. While one or two existing works outperform our proposed method on the LIVE-Qualcomm and CVD2014, this is likely due to the artificial nature of degradations in these datasets that our feature extraction network is not trained on. We also show that our proposed method outperforms the current state-of-the-art on KonVid-150k-B, the set of 1,596 accurately labeled videos that are part of our proposed dataset. Additionally, by training our method on the entirety of the proposed noisily annotated dataset, we can improve the cross-dataset test performance on KoNViD-1k, LIVE-Qualcomm, and LIVE-VQC, even beating within-dataset performances on KoNViD-1k and LIVE-VQC. CVD2014 appears to be a tough challenge for our approach, both when trained in within-dataset and cross-dataset scenarios.
	
	\section*{Acknowledgment}
	Funded by the Deutsche Forschungsgemeinschaft (DFG, German Research Foundation) -- Project-ID 251654672 -- TRR 161 (Project A05).
	

\bibliographystyle{IEEEtran}

\begin{thebibliography}{10}
	\providecommand{\url}[1]{#1}
	\csname url@samestyle\endcsname
	\providecommand{\newblock}{\relax}
	\providecommand{\bibinfo}[2]{#2}
	\providecommand{\BIBentrySTDinterwordspacing}{\spaceskip=0pt\relax}
	\providecommand{\BIBentryALTinterwordstretchfactor}{4}
	\providecommand{\BIBentryALTinterwordspacing}{\spaceskip=\fontdimen2\font plus
		\BIBentryALTinterwordstretchfactor\fontdimen3\font minus
		\fontdimen4\font\relax}
	\providecommand{\BIBforeignlanguage}[2]{{%
			\expandafter\ifx\csname l@#1\endcsname\relax
			\typeout{** WARNING: IEEEtran.bst: No hyphenation pattern has been}%
			\typeout{** loaded for the language `#1'. Using the pattern for}%
			\typeout{** the default language instead.}%
			\else
			\language=\csname l@#1\endcsname
			\fi
			#2}}
	\providecommand{\BIBdecl}{\relax}
	\BIBdecl
	
	\bibitem{wyzowl}
	Wyzowl, ``{Wyzowl State of Video Marketing Statistics 2019},''
	\url{https://info.wyzowl.com/state-of-video-marketing-2019-report}, 2019,
	[Online; accessed 15-November-2019].
	
	\bibitem{buffer}
	Buffer, ``{State Of Social 2019 Report},''
	\url{https://buffer.com/state-of-social-2019}, 2019, [Online; accessed
	15-November-2019].
	
	\bibitem{deloitte}
	K.~Westcott, J.~Loucks, K.~Downs, and J.~Watson, ``Digital {M}edia {T}rends
	{S}urvey, 12th edition,'' 2018.
	
	\bibitem{cisco2018cisco}
	{Cisco, VNI}, ``Cisco visual networking index: Forecast and trends,
	2017--2022,'' \emph{White Paper}, vol.~1, 2018.
	
	\bibitem{youtube}
	C.~Goodrow, ``You know what’s cool? a billion hours,''
	\url{https://youtube.googleblog.com/2017/02/you-know-whats-cool-billion-hours.html},
	2017, [Online; accessed 15-November-2019].
	
	\bibitem{mckechnie1977simulation}
	G.~E. McKechnie, ``Simulation techniques in environmental psychology,'' in
	\emph{Perspectives on environment and behavior}.\hskip 1em plus 0.5em minus
	0.4em\relax Springer, 1977, pp. 169--189.
	
	\bibitem{ang2013data}
	C.~S. Ang, A.~Bobrowicz, D.~J. Schiano, and B.~Nardi, ``Data in the wild: Some
	reflections,'' \emph{interactions}, vol.~20, no.~2, pp. 39--43, 2013.
	
	\bibitem{argyropoulos2011no}
	S.~Argyropoulos, A.~Raake, M.-N. Garcia, and P.~List, ``No-reference video
	quality assessment for {SD} and {HD} {H}. 264/{AVC} sequences based on
	continuous estimates of packet loss visibility,'' in \emph{2011 Third
		International Workshop on Quality of Multimedia Experience}.\hskip 1em plus
	0.5em minus 0.4em\relax IEEE, 2011, pp. 31--36.
	
	\bibitem{chen2011prediction}
	Z.~Chen and D.~Wu, ``Prediction of transmission distortion for wireless video
	communication: Analysis,'' \emph{IEEE Transactions on Image Processing},
	vol.~21, no.~3, pp. 1123--1137, 2011.
	
	\bibitem{valenzise2011no}
	G.~Valenzise, S.~Magni, M.~Tagliasacchi, and S.~Tubaro, ``No-reference pixel
	video quality monitoring of channel-induced distortion,'' \emph{IEEE
		Transactions on Circuits and Systems for Video Technology}, vol.~22, no.~4,
	pp. 605--618, 2011.
	
	\bibitem{saad2014blind}
	M.~A. Saad, A.~C. Bovik, and C.~Charrier, ``Blind prediction of natural video
	quality,'' \emph{IEEE Transactions on Image Processing}, vol.~23, no.~3, pp.
	1352--1365, 2014.
	
	\bibitem{pandremmenou2015no}
	K.~Pandremmenou, M.~Shahid, L.~P. Kondi, and B.~L{\"o}vstr{\"o}m, ``A
	no-reference bitstream-based perceptual model for video quality estimation of
	videos affected by coding artifacts and packet losses,'' in \emph{Human
		Vision and Electronic Imaging XX}, vol. 9394.\hskip 1em plus 0.5em minus
	0.4em\relax International Society for Optics and Photonics, 2015, p. 93941F.
	
	\bibitem{keimel2011design}
	C.~Keimel, J.~Habigt, M.~Klimpke, and K.~Diepold, ``Design of no-reference
	video quality metrics with multiway partial least squares regression,'' in
	\emph{2011 Third International Workshop on Quality of Multimedia
		Experience}.\hskip 1em plus 0.5em minus 0.4em\relax IEEE, 2011, pp. 49--54.
	
	\bibitem{zhu2014no}
	K.~Zhu, C.~Li, V.~Asari, and D.~Saupe, ``No-reference video quality assessment
	based on artifact measurement and statistical analysis,'' \emph{IEEE
		Transactions on Circuits and Systems for Video Technology}, vol.~25, no.~4,
	pp. 533--546, 2014.
	
	\bibitem{sogaard2015no}
	J.~S{\o}gaard, S.~Forchhammer, and J.~Korhonen, ``No-reference video quality
	assessment using codec analysis,'' \emph{IEEE Transactions on Circuits and
		Systems for Video Technology}, vol.~25, no.~10, pp. 1637--1650, 2015.
	
	\bibitem{mittal2015completely}
	A.~Mittal, M.~A. Saad, and A.~C. Bovik, ``A completely blind video integrity
	oracle,'' \emph{IEEE Transactions on Image Processing}, vol.~25, no.~1, pp.
	289--300, 2015.
	
	\bibitem{vega2017predictive}
	M.~T. Vega, D.~C. Mocanu, S.~Stavrou, and A.~Liotta, ``Predictive no-reference
	assessment of video quality,'' \emph{Signal Processing: Image Communication},
	vol.~52, pp. 20--32, 2017.
	
	\bibitem{korhonen2018learning}
	J.~Korhonen, ``Learning-based prediction of packet loss artifact visibility in
	networked video,'' in \emph{2018 Tenth International Conference on Quality of
		Multimedia Experience}.\hskip 1em plus 0.5em minus 0.4em\relax IEEE, 2018,
	pp. 1--6.
	
	\bibitem{korhonen2019hierarchical}
	J.~Korhnen, ``Two-level approach for no-reference consumer video quality
	assessment,'' \emph{IEEE Transactions on Image Processing}, vol.~28, no.~12,
	pp. 5923--5938, 2019.
	
	\bibitem{konvid}
	V.~Hosu, F.~Hahn, M.~Jenadeleh, H.~Lin, H.~Men, T.~Szir{\'a}nyi, S.~Li, and
	D.~Saupe, ``The {Konstanz Natural Video Database KoNViD-1k},'' in \emph{9th
		International Conference on Quality of Multimedia Experience}, 2017.
	
	\bibitem{CVD2014}
	M.~Nuutinen, T.~Virtanen, M.~Vaahteranoksa, T.~Vuori, P.~Oittinen, and
	J.~H{\"a}kkinen, ``{CVD2014}— a database for evaluating no-reference video
	quality assessment algorithms,'' \emph{IEEE Transactions on Image
		Processing}, vol.~25, no.~7, pp. 3073--3086, 2016.
	
	\bibitem{ghadiyaram2017capture}
	D.~Ghadiyaram, J.~Pan, A.~C. Bovik, A.~K. Moorthy, P.~Panda, and K.-C. Yang,
	``In-capture mobile video distortions: A study of subjective behavior and
	objective algorithms,'' \emph{IEEE Transactions on Circuits and Systems for
		Video Technology}, 2017.
	
	\bibitem{sinno2019large}
	Z.~Sinno and A.~C. Bovik, ``Large-scale study of perceptual video quality,''
	\emph{IEEE Transactions on Image Processing}, vol.~28, no.~2, pp. 612--627,
	2019.
	
	\bibitem{gao2018blind}
	F.~Gao, J.~Yu, S.~Zhu, Q.~Huang, and Q.~Tian, ``Blind image quality prediction
	by exploiting multi-level deep representations,'' \emph{Pattern Recognition},
	vol.~81, pp. 432--442, 2018.
	
	\bibitem{zhang2018unreasonable}
	R.~Zhang, P.~Isola, A.~A. Efros, E.~Shechtman, and O.~Wang, ``The unreasonable
	effectiveness of deep features as a perceptual metric,'' in \emph{IEEE
		Conference on Computer Vision and Pattern Recognition}, 2018, pp. 586--595.
	
	\bibitem{hosu2019effective}
	V.~Hosu, B.~Goldlucke, and D.~Saupe, ``Effective aesthetics prediction with
	multi-level spatially pooled features,'' in \emph{IEEE Conference on Computer
		Vision and Pattern Recognition}, 2019, pp. 9375--9383.
	
	\bibitem{szegedy2017inception}
	C.~Szegedy, S.~Ioffe, V.~Vanhoucke, and A.~A. Alemi, ``Inception-v4,
	inception-resnet and the impact of residual connections on learning,'' in
	\emph{Thirty-First AAAI Conference on Artificial Intelligence}, 2017.
	
	\bibitem{ben2007analysis}
	S.~Ben-David, J.~Blitzer, K.~Crammer, and F.~Pereira, ``Analysis of
	representations for domain adaptation,'' in \emph{Advances in Neural
		Information Processing Systems}, 2007, pp. 137--144.
	
	\bibitem{de2009subjective}
	F.~De~Simone, M.~Naccari, M.~Tagliasacchi, F.~Dufaux, S.~Tubaro, and
	T.~Ebrahimi, ``Subjective assessment of {H}. 264/{AVC} video sequences
	transmitted over a noisy channel,'' in \emph{2009 International Workshop on
		Quality of Multimedia Experience}.\hskip 1em plus 0.5em minus 0.4em\relax
	IEEE, 2009, pp. 204--209.
	
	\bibitem{de2010h}
	F.~De~Simone, M.~Tagliasacchi, M.~Naccari, S.~Tubaro, and T.~Ebrahimi, ``A {H}.
	264/{AVC} video database for the evaluation of quality metrics,'' in
	\emph{2010 IEEE International Conference on Acoustics Speech and Signal
		Processing}.\hskip 1em plus 0.5em minus 0.4em\relax IEEE, 2010, pp.
	2430--2433.
	
	\bibitem{seshadrinathan2010study}
	K.~Seshadrinathan, R.~Soundararajan, A.~C. Bovik, and L.~K. Cormack, ``Study of
	subjective and objective quality assessment of video,'' \emph{IEEE
		Transactions on Image Processing}, vol.~19, no.~6, pp. 1427--1441, 2010.
	
	\bibitem{seshadrinathan2010subjective}
	------, ``A subjective study to evaluate video quality assessment algorithms,''
	in \emph{Human Vision and Electronic Imaging XV}, vol. 7527.\hskip 1em plus
	0.5em minus 0.4em\relax International Society for Optics and Photonics, 2010,
	p. 75270H.
	
	\bibitem{larson2010most}
	E.~C. Larson and D.~M. Chandler, ``Most apparent distortion: full-reference
	image quality assessment and the role of strategy,'' \emph{Journal of
		Electronic Imaging}, vol.~19, no.~1, p. 011006, 2010.
	
	\bibitem{video2010report}
	{Video Quality Experts Group}, ``Report on the validation of video quality
	models for high definition video content,'' \emph{http://www. its. bldrdoc.
		gov/media/4212/vqeg\_hdtv\_final\_report\_version\_2. 0. zip}, 2010.
	
	\bibitem{zhang2011ivp}
	F.~Zhang, S.~Li, L.~Ma, Y.~C. Wong, and K.~N. Ngan, ``{IVP} subjective quality
	video database,'' \emph{The Chinese University of Hong Kong, http://ivp. ee.
		cuhk. edu. hk/research/database/subjective}, 2011.
	
	\bibitem{saupe2016crowd}
	D.~Saupe, F.~Hahn, V.~Hosu, I.~Zingman, M.~Rana, and S.~Li, ``Crowd workers
	proven useful: A comparative study of subjective video quality assessment,''
	in \emph{QoMEX 2016: International Conference on Quality of Multimedia
		Experience}, 2016.
	
	\bibitem{rolnick2017deep}
	D.~Rolnick, A.~Veit, S.~Belongie, and N.~Shavit, ``Deep learning is robust to
	massive label noise,'' \emph{arXiv preprint arXiv:1705.10694}, 2017.
	
	\bibitem{varga2020multi}
	D.~Varga, ``Multi-pooled inception features for no-reference video quality
	assessment.'' in \emph{VISIGRAPP (4: VISAPP)}, 2020, pp. 338--347.
	
	\bibitem{srivastava2003advances}
	A.~Srivastava, A.~B. Lee, E.~P. Simoncelli, and S.-C. Zhu, ``On advances in
	statistical modeling of natural images,'' \emph{Journal of Mathematical
		Imaging and Vision}, vol.~18, no.~1, pp. 17--33, 2003.
	
	\bibitem{mittal2012making}
	A.~Mittal, R.~Soundararajan, and A.~C. Bovik, ``Making a “completely blind”
	image quality analyzer,'' \emph{IEEE Signal Processing Letters}, vol.~20,
	no.~3, pp. 209--212, 2012.
	
	\bibitem{mittal2012no}
	A.~Mittal, A.~K. Moorthy, and A.~C. Bovik, ``No-reference image quality
	assessment in the spatial domain,'' \emph{IEEE Transactions on Image
		Processing}, vol.~21, no.~12, pp. 4695--4708, 2012.
	
	\bibitem{xu2014no}
	J.~Xu, P.~Ye, Y.~Liu, and D.~Doermann, ``No-reference video quality assessment
	via feature learning,'' in \emph{2014 IEEE International Conference on Image
		Processing}.\hskip 1em plus 0.5em minus 0.4em\relax IEEE, 2014, pp. 491--495.
	
	\bibitem{kundu2017no}
	D.~Kundu, D.~Ghadiyaram, A.~C. Bovik, and B.~L. Evans, ``No-reference quality
	assessment of tone-mapped {HDR} pictures,'' \emph{IEEE Transactions on Image
		Processing}, vol.~26, no.~6, pp. 2957--2971, 2017.
	
	\bibitem{li2016no}
	Y.~Li, L.-M. Po, C.-H. Cheung, X.~Xu, L.~Feng, F.~Yuan, and K.-W. Cheung,
	``No-reference video quality assessment with 3{D} shearlet transform and
	convolutional neural networks.'' \emph{IEEE Transaction Circuits System Video
		Technology}, vol.~26, no.~6, pp. 1044--1057, 2016.
	
	\bibitem{wang2018come}
	C.~Wang, L.~Su, and W.~Zhang, ``{COME} for no-reference video quality
	assessment,'' in \emph{2018 IEEE Conference on Multimedia Information
		Processing and Retrieval}.\hskip 1em plus 0.5em minus 0.4em\relax IEEE, 2018,
	pp. 232--237.
	
	\bibitem{you2019deep}
	J.~You and J.~Korhonen, ``Deep neural networks for no-reference video quality
	assessment,'' in \emph{2019 IEEE International Conference on Image Processing
		(ICIP)}.\hskip 1em plus 0.5em minus 0.4em\relax IEEE, 2019, pp. 2349--2353.
	
	\bibitem{varga2019no}
	D.~Varga, ``No-reference video quality assessment based on the temporal pooling
	of deep features,'' \emph{Neural Processing Letters}, vol.~50, no.~3, pp.
	2595--2608, 2019.
	
	\bibitem{varga2019no2}
	D.~Varga and T.~Szir{\'{a}}nyi, ``No-reference video quality assessment via
	pretrained {CNN} and {LSTM} networks,'' \emph{Signal, Image and Video
		Processing}, vol.~13, no.~8, pp. 1569--1576, 2019.
	
	\bibitem{gotz2020critical}
	F.~G{\"o}tz-Hahn, V.~Hosu, and D.~Saupe, ``Critical analysis on the
	reproducibility of visual quality assessment using deep features,''
	\emph{arXiv preprint arXiv:2009.05369}, 2020.
	
	\bibitem{hosu2020koniq}
	V.~Hosu, H.~Lin, T.~Sziranyi, and D.~Saupe, ``Kon{IQ}-10k: {A}n ecologically
	valid database for deep learning of blind image quality assessment,''
	\emph{IEEE Transactions on Image Processing}, vol.~29, pp. 4041--4056, 2020.
	
	\bibitem{tutorial2004objective}
	{ITU-T}, ``Objective perceptual assessment of video quality: Full reference
	television,'' Tutorial, ITU-T Telecommunication Standardization Bureau, 2004.
	
	\bibitem{hossfeld2011sos}
	T.~Ho{\ss}feld, R.~Schatz, and S.~Egger, ``{SOS}: The {MOS} is not enough!'' in
	\emph{Third International Workshop on Quality of Multimedia Experience},
	2011, pp. 131--136.
	
	\bibitem{janowski2015accuracy}
	L.~Janowski and M.~Pinson, ``The accuracy of subjects in a quality experiment:
	A theoretical subject model,'' \emph{IEEE Transactions on Multimedia},
	vol.~17, no.~12, pp. 2210--2224, 2015.
	
	\bibitem{lin2020deepfl}
	H.~Lin, V.~Hosu, and D.~Saupe, ``Deepfl-iqa: Weak supervision for deep iqa
	feature learning,'' \emph{arXiv preprint arXiv:2001.08113}, 2020.
	
	\bibitem{hermans2013training}
	M.~Hermans and B.~Schrauwen, ``Training and analysing deep recurrent neural
	networks,'' in \emph{Advances in Neural Information Processing Systems},
	2013, pp. 190--198.
	
\end{thebibliography}
\end{document}